\documentclass{article}
\newcommand{\preprinttitle}{NEAR: Neighborhood-Aware Residuals for Memory Retrieval}
\PassOptionsToPackage{table}{xcolor}
\usepackage{fix-cm}
\usepackage[T1]{fontenc}
\usepackage{near_preprint,times}
\usepackage{amsmath,amssymb,amsthm,graphicx,booktabs,array,xcolor,placeins}
\usepackage{hyperref,url}
\usepackage{caption}
\usepackage{float}
\usepackage{needspace}
\usepackage{booktabs,array,tabularx,xcolor,colortbl}
\definecolor{paperLink}{RGB}{62,82,112}
\definecolor{nearTableTint}{RGB}{232,240,247}
\colorlet{neutralTableGray}{nearTableTint}
\definecolor{nearGroupTint}{RGB}{245,245,245}
\newlength{\nearTableEdgePadding}
\newcommand{\nearUnifiedTableStyle}{%
  \arrayrulecolor{black!75}%
  \setlength{\arrayrulewidth}{0.35pt}%
  \setlength{\heavyrulewidth}{0.8pt}%
  \setlength{\lightrulewidth}{0.35pt}%
  \setlength{\cmidrulewidth}{0.3pt}%
  \setlength{\aboverulesep}{1.2pt}%
  \setlength{\belowrulesep}{1.2pt}}
\newcommand{\nearTableRules}{%
  \setlength{\arrayrulewidth}{0.35pt}%
  \arrayrulecolor{black!75}}

\newcommand{\neutralMainTableStyle}{%
  \fontsize{7.6}{8.2}\selectfont
  \setlength{\tabcolsep}{3pt}%
  \renewcommand{\arraystretch}{0.92}%
  \setlength{\aboverulesep}{1.2pt}%
  \setlength{\belowrulesep}{1.2pt}}

\newcommand{\neutralSummaryTableStyle}{%
  \fontsize{8.2}{9.5}\selectfont
  \setlength{\tabcolsep}{2.5pt}%
  \renewcommand{\arraystretch}{0.98}}

\hypersetup{colorlinks=true,linkcolor=paperLink,citecolor=paperLink,urlcolor=paperLink}
\colorlet{oursblue}{nearTableTint}
\definecolor{groupgray}{RGB}{239,240,241}
\definecolor{gaingreen}{RGB}{36,126,69}

\newcommand{\ours}{\textsc{NEAR}}
\newcommand{\unit}{\operatorname{unit}}

\newcommand{\R}{\mathbb R}

\newcommand{\sdsmall}[1]{\,{\fontsize{5.5}{6.1}\selectfont$\pm$#1}}

\title{NEAR: Neighborhood-Aware Residuals\\for Memory Retrieval}
\author{%
\begin{minipage}[t]{\dimexpr\textwidth-2\tabcolsep\relax}
\raggedright
Han Luo\textsuperscript{1}\quad
Zhan Zhuang\textsuperscript{2}\quad
Xu He\textsuperscript{2}\quad Jujia Zhao\textsuperscript{3}\\
Yu Li\textsuperscript{1}\quad
Suqin Yuan\textsuperscript{4}\quad
Dong Li\textsuperscript{2}\quad Lei Feng\textsuperscript{1}\thanks{Corresponding author.}\\[3pt]
{\normalfont\normalsize
\textsuperscript{1}Southeast University\quad
\textsuperscript{2}MemoraX AI\quad
\textsuperscript{3}Leiden University\quad
\textsuperscript{4}University of Sydney}
\end{minipage}}
\iclrfinalcopy
\hypersetup{
  pdftitle={NEAR: Neighborhood-Aware Residuals for Memory Retrieval},
  pdfauthor={Han Luo, Zhan Zhuang, Xu He, Jujia Zhao, Yu Li, Suqin Yuan, Dong Li, Lei Feng},
  pdfsubject={Memory Retrieval},
  pdfkeywords={memory retrieval, dense retrieval, frozen embeddings, neighborhood-aware representations, approximate nearest-neighbor search}}
\begin{document}
\raggedbottom
\makeatletter\setlength{\@fptop}{0pt}\makeatother
\maketitle
\renewcommand{\headrulewidth}{0.4pt}
\vspace{-20pt}

\begin{abstract}
Memory retrieval aims to identify evidence from stored records to answer a query.
Stored records with different factual details may share topics, entities, or events.
Such shared contents would make records without the required evidence more likely to receive high similarity scores and outrank the correct evidence.
In this paper, we use the information shared by these records to construct local references for query--memory matching.
We propose \textbf{NE}ighborhood-\textbf{A}ware \textbf{R}esiduals (NEAR), a training-free framework that uses these references to construct residual representations that complement the original embeddings.
Specifically, each memory derives a reference direction from its neighbors, and each query obtains its reference from a precomputed table.
Using these references, NEAR constructs projection residuals for both queries and memories.
These residuals are combined with the original embeddings for a single approximate nearest-neighbor search over a fixed memory index.
Experiments on eight long-context retrieval benchmarks show consistent improvements across nine embedding models.
Downstream evaluations show that NEAR improves question-answering performance.
\end{abstract}

\section{Introduction}
\label{sec:introduction}

Memory retrieval enables intelligent assistants to find evidence in past conversations and stored documents to answer current queries \citep{locomo,longmem}.
A common approach is embedding-based retrieval, which encodes queries and memories separately and ranks memories by vector similarity \citep{cde}.
Memory embeddings can be computed and indexed in advance, allowing new queries to search the memory bank without re-encoding its records \citep{ashok2026sharpening}.
This design separates offline memory processing from online query retrieval.

\begin{figure}[t]
\centering
\includegraphics[width=\linewidth]{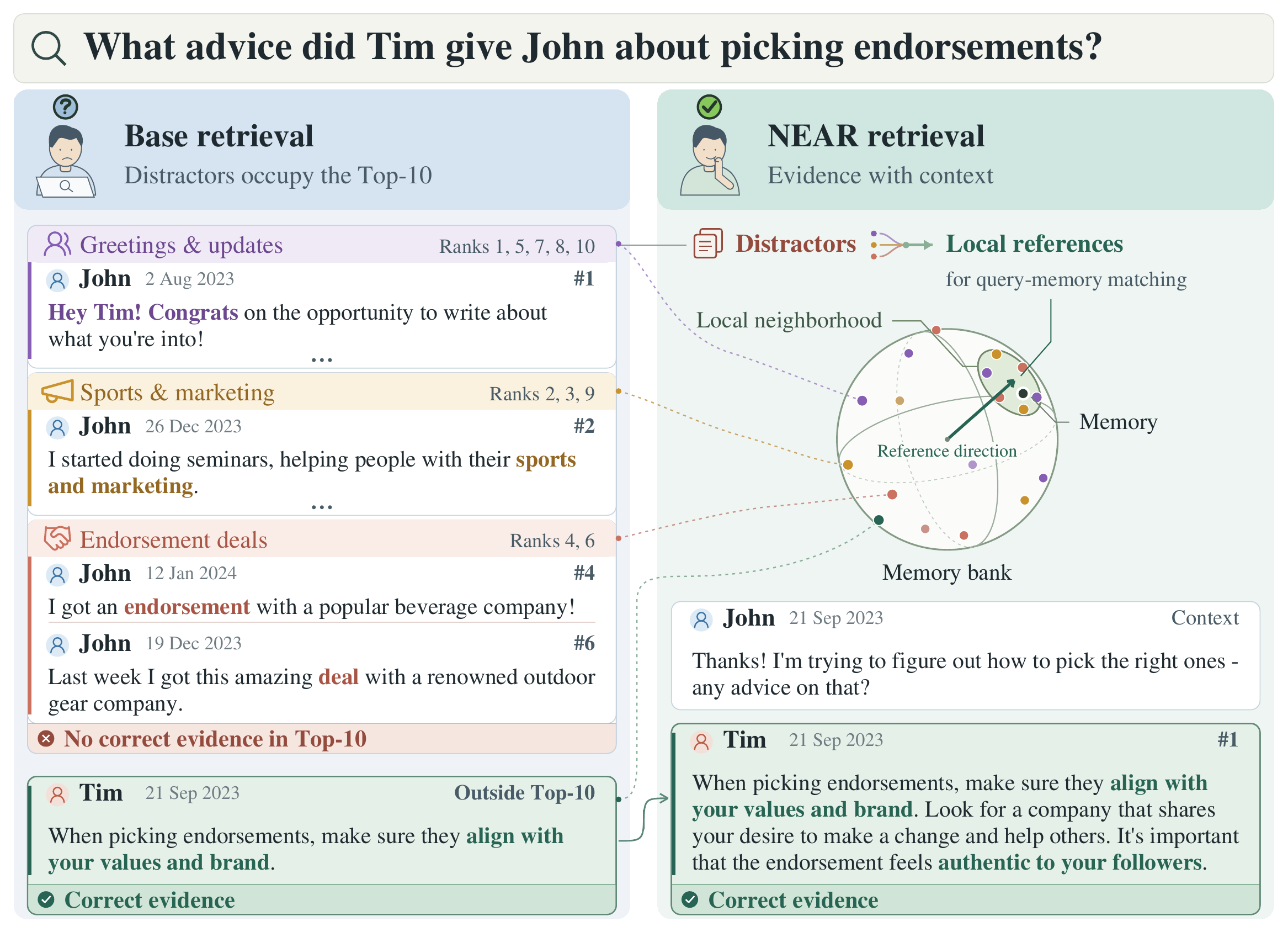}
\caption{\textbf{Turning distractors into local references.} In this LoCoMo example (Qwen3-Embedding-4B), Base's Top-10 contains records about the same people and related topics but misses Tim's advice on choosing endorsements. However, these shared contents also provide useful information for distinguishing the candidates. NEAR uses this information to construct local references for query--memory matching and ranks the missed evidence first.}
\label{fig:motivation}
\label{fig:case-study}
\end{figure}

In a memory bank, records with different factual details may share topics, entities, or events \citep{assomem2026}.
When ranking relies only on query--memory similarity, such overlap would make records that do not provide the required evidence more likely to outrank correct evidence \citep{ashok2026sharpening}.
However, these similar records are not only competing candidates.
We argue that they also provide a local neighborhood that can serve as a reference for each memory.
We turn this potential source of confusion into useful matching information by constructing memory-specific local references from the relationships among similar memories, as illustrated in Figure~\ref{fig:motivation}.

In this paper, we propose \textbf{NE}ighborhood-\textbf{A}ware \textbf{R}esiduals (NEAR), a training-free framework that uses these local references for query--memory matching.
Figure~\ref{fig:method} gives an overview of NEAR.
For each memory, NEAR builds a reference from its nearest neighbors and removes the embedding component along this direction.
A small precomputed table provides references for new queries without an initial search over the memory bank.
The same projection yields a residual representation for each query.
NEAR combines the original and residual representations so that the final matching score is a single inner product.
Offline indexing enables retrieval through a single approximate nearest-neighbor (ANN) search over the fixed memory index.

We evaluate NEAR on eight long-context retrieval benchmarks using nine frozen embedding models.
Across all nine models, NEAR improves mean Recall@10 and NDCG@10 over the original embeddings and the evaluated same-backbone baselines.
Downstream evaluations further show that NEAR improves question-answering accuracy by 9.6 percentage points on LoCoMo with Qwen3-Embedding-4B as the retrieval model.
Controlled comparisons show that local reference construction and memory--reference correspondence contribute to the retrieval gains.
Additional analyses characterize parameter sensitivity and the trade-offs between retrieval quality, storage, and query latency, including the benefits of index compression.

Our main contributions are as follows:
\begingroup
\settowidth{\leftmargini}{Ou\labelitemi}
\addtolength{\leftmargini}{\labelsep}
\begin{itemize}
    \item We propose NEAR, a training-free method that constructs memory-specific local references from shared contents and uses the resulting residual representations to complement the original embeddings for query--memory matching.
    \item We construct a compact query-reference table and a fixed memory index, enabling new queries to retrieve memories through a single ANN search without preliminary candidate retrieval.
    \item We demonstrate improvements in mean Recall@10 and NDCG@10 across eight retrieval benchmarks and nine embedding models, together with downstream question-answering gains.
\end{itemize}
\endgroup

\begingroup
\setlength{\parskip}{3pt}
\section{Related work}
\noindent\textbf{Long-term memory and retrieval.}
Long-term memory research explores how to organize, update, and retrieve past information to support question answering over ongoing interactions.
A-Mem organizes structured memory notes into a dynamically linked network, while MemoryOS manages memories at different time scales through a hierarchical architecture \citep{amem2025,memoryos2025}.
LightMem reduces memory-processing costs through information compression, topic-based consolidation, and offline updates \citep{lightmem2026}.
For retrieval, AssoMem uses an associative memory graph and combines relevance, importance, and temporal signals for ranking \citep{assomem2026}.
EvoEmbedding instead learns context-dependent retrieval representations using a continuously updated latent memory \citep{evoembedding2026}.
We focus on scoring queries against a given memory bank, without changing its content or updating the pretrained encoder.

\noindent\textbf{Embedding adaptation and similarity calibration.}
Contextual embedding methods incorporate surrounding text or other documents during encoding \citep{cde,insent2025}.
Existing embeddings can be adapted for retrieval through supervised mappings or training-free corrections.
Search-Adaptor learns transformations of pretrained embeddings from relevance supervision while keeping the encoder fixed \citep{searchadaptor}.
Training-free methods use neighborhoods or reference collections, but differ in whether they adjust scores or vectors.
LC corrects similarity scores using local affinity estimates to reduce hubness \citep{lc}.
For cross-modal retrieval, NNN estimates matching biases from a reference database, while DBSN uses query and target reference banks to calibrate score distributions \citep{nnn2024,dbsn}.
Rather than correcting scores directly, InvGC transforms embedding vectors through inverse graph convolution with global or local reference sets \citep{invgc}.

\noindent\textbf{Residual representations and reference-based matching.}
Residual-based methods compare or encode representations after removing a reference component.
Local residual similarity constructs neighborhood-derived residuals for image re-ranking \citep{lrs}.
PRQ-KMeans instead recursively removes selected-centroid projections to construct semantic IDs for generative retrieval and recommendation \citep{prq}.
For first-stage memory retrieval, memory-specific references can be computed offline, but a new query also needs a reference before any candidates have been retrieved.
NEAR addresses this requirement by combining each memory's neighborhood reference with a precomputed query-reference mapping.
It retains original similarity alongside residual matching and expresses their combined score as an inner product over a fixed memory index, enabling a single ANN search without preliminary candidate retrieval.

\begin{figure}[t]
\centering\includegraphics[width=\linewidth]{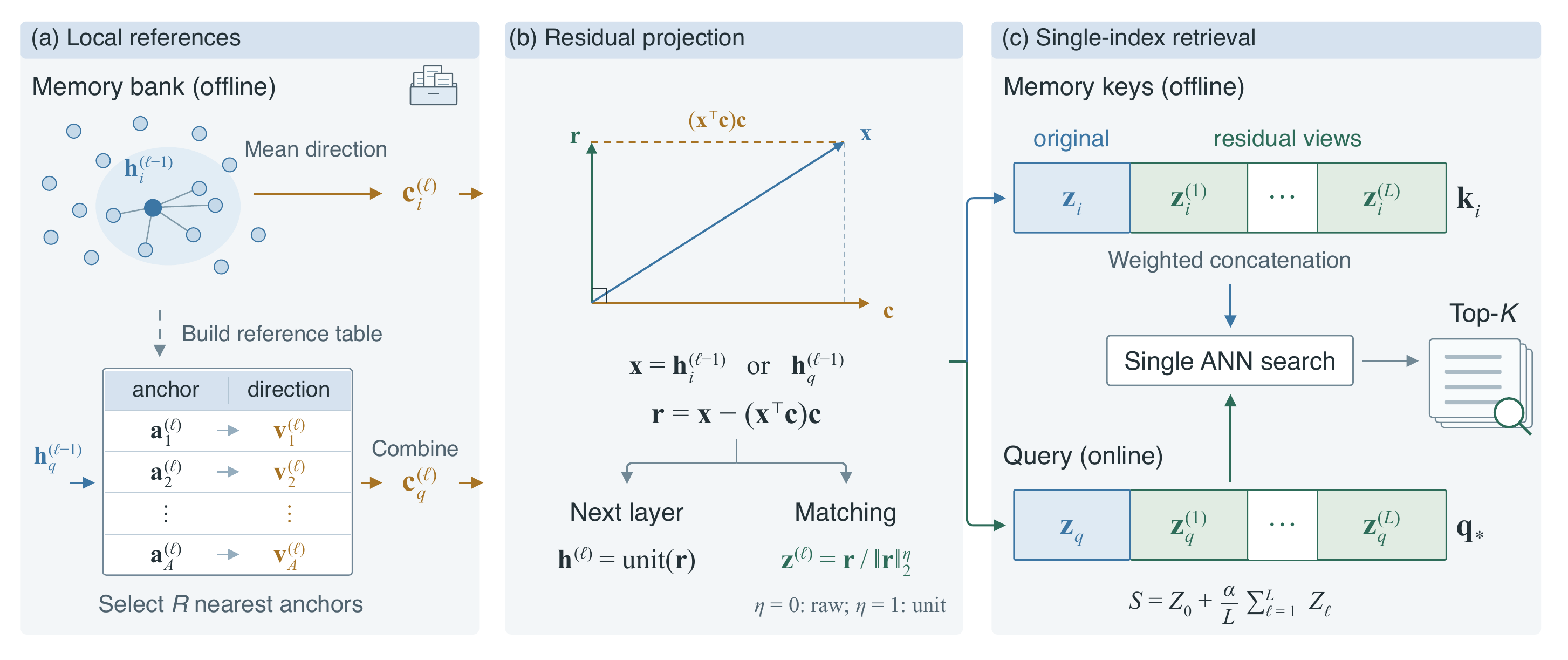}
\caption{\textbf{Overview of NEAR.} Memory neighborhoods yield local references, while a table supplies query references. Original and residual views form static keys for one ANN search.}
\label{fig:method}
\end{figure}

\section{Method}
\label{sec:method}
Given a text query $q$ and a bank of memory records $\mathcal{M}=\{m_i\}_{i=1}^{N}$, we use a frozen encoder to obtain unit-length embeddings $\mathbf{z}_q,\mathbf{z}_i\in\R^d$.
Original retrieval ranks memories by their cosine similarity $\mathbf{z}_q^{\top}\mathbf{z}_i$.
NEAR complements this score with residual representations constructed from memory neighborhoods.
It builds memory representations and a query-reference table offline. A new query uses the table to form its residuals before a single ANN search.

\subsection{Neighborhood-Based Memory Representations}
Initialize $\mathbf{h}_i^{(0)}=\mathbf{z}_i$.
At layer $\ell$, let $\mathcal{N}_i^{(\ell)}$ contain the $\min(k,N-1)$ nearest other memories to $\mathbf{h}_i^{(\ell-1)}$ by cosine similarity.
Their mean direction defines the local reference
\begin{gather*}
\mathbf{c}_i^{(\ell)}=\unit\left(
\frac{1}{|\mathcal{N}_i^{(\ell)}|}
\sum\nolimits_{j\in\mathcal{N}_i^{(\ell)}}\mathbf{h}_j^{(\ell-1)}
\right),
\end{gather*}
where $\unit(\mathbf{v})=\mathbf{v}/\|\mathbf{v}\|_2$ for nonzero vectors.
Removing the projection onto this direction gives the residual
\begin{gather}
\mathbf{r}_i^{(\ell)}
=\mathbf{h}_i^{(\ell-1)}
-\left((\mathbf{h}_i^{(\ell-1)})^{\top}\mathbf{c}_i^{(\ell)}\right)\mathbf{c}_i^{(\ell)}.
\label{eq:operator}
\end{gather}

For subsequent layers, we repeat this procedure using the normalized residuals $\mathbf{h}_i^{(\ell)}=\unit(\mathbf{r}_i^{(\ell)})$.
The matching representation either retains the residual length or is normalized:
\begin{gather}
\mathbf{z}_i^{(\ell)}=
\begin{cases}
\mathbf{r}_i^{(\ell)}, & \eta=0,\\
\unit(\mathbf{r}_i^{(\ell)}), & \eta=1,
\end{cases}
\label{eq:residualstate}
\end{gather}
where $\eta$ is selected on development data and shared by memories and queries.
We retain the original embeddings and the matching representations from all $L$ layers for score fusion.

\subsection{Precomputed Query References}
At each layer, NEAR precomputes a small table that maps queries to local reference directions.
Spherical K-means \citep{sphericalkmeans} clusters the input memory states, with cluster centers serving as anchors $\mathbf{a}_h^{(\ell)}$.
Each anchor stores $\mathbf{v}_h^{(\ell)}$, the normalized mean reference direction of the memories in its cluster.

Initialize the query state $\mathbf{h}_q^{(0)}=\mathbf{z}_q$.
At layer $\ell$, the query selects the $R=\min(R_{\max},A)$ most similar anchors from the table's $A$ entries, with index set $\mathcal{H}_q^{(\ell)}$.
Similarity-based softmax weights combine their stored directions into a query reference:
\begin{gather*}
p_h^{(\ell)}(q)
=\frac{\exp\{\beta(\mathbf{h}_q^{(\ell-1)})^{\top}\mathbf{a}_h^{(\ell)}\}}
{\sum\nolimits_{j\in\mathcal{H}_q^{(\ell)}}\exp\{\beta(\mathbf{h}_q^{(\ell-1)})^{\top}\mathbf{a}_j^{(\ell)}\}},
\qquad
\mathbf{c}_q^{(\ell)}
=\unit\left(\sum\nolimits_{h\in\mathcal{H}_q^{(\ell)}}p_h^{(\ell)}(q)\mathbf{v}_h^{(\ell)}\right),
\end{gather*}
where $\beta$ controls weight concentration.
The same projection and normalization rules yield the query matching representation $\mathbf{z}_q^{(\ell)}$ and next-layer state $\mathbf{h}_q^{(\ell)}$, and Appendix~\ref{app:population} lists the parameter settings.

\subsection{Score Fusion and Single-Index Retrieval}
\label{sec:index}
Let $\mathbf{z}_x^{(0)}=\mathbf{z}_x$ denote the original embedding and $s_\ell(q,m_i)=(\mathbf{z}_q^{(\ell)})^{\top}\mathbf{z}_i^{(\ell)}$ the matching score at layer $\ell$.
NEAR combines original and residual matching as
\begin{gather}
S(q,m_i)
=\frac{s_0(q,m_i)}{\tau_0}
+\frac{\alpha}{L}\sum\nolimits_{\ell=1}^{L}
\frac{s_\ell(q,m_i)}{\tau_\ell},
\label{eq:score}
\end{gather}
where $\alpha\ge0$ controls the total residual weight independently of depth.
Each scale $\tau_\ell$ is computed offline from the standard deviations of memory--memory inner products and shared across queries for the same bank (Appendix~\ref{app:proof}).
We write $Z_\ell=s_\ell/\tau_\ell$ for the scaled score.

The fused score can be evaluated as one inner product by concatenating representations into a static memory key and a query vector:
\begin{align}
\mathbf{k}_i
&=\left[\mathbf{z}_i;\ \sqrt{\alpha/L}\,\mathbf{z}_i^{(1)};\ \ldots;\ \sqrt{\alpha/L}\,\mathbf{z}_i^{(L)}\right],\nonumber\\
\mathbf{q}_*
&=\left[\mathbf{z}_q/\tau_0;\ \frac{\sqrt{\alpha/L}}{\tau_1}\mathbf{z}_q^{(1)};\ \ldots;\ \frac{\sqrt{\alpha/L}}{\tau_L}\mathbf{z}_q^{(L)}\right],
\label{eq:compile}
\end{align}
where semicolons denote concatenation.
Their inner product equals the fused score exactly: $\mathbf{q}_*^{\top}\mathbf{k}_i=S(q,m_i)$.
Memory keys can therefore be constructed and indexed offline.
A new query forms $\mathbf{q}_*$ using the reference tables and performs a single inner-product ANN search to retrieve memories.

\par\endgroup

\begingroup
\setlength{\parskip}{5pt}
\section{Experiments}
\subsection{Experimental Setup}
\label{sec:experimental-setup}
\noindent\textbf{Benchmarks.}
We evaluate retrieval on eight benchmarks: LoCoMo \citep{locomo}, LongMemEval \citep{longmem}, and REALTALK \citep{realtalk} for conversational memory, and ESG-Reports \citep{vidorev2}, QASPER \citep{qasper}, PeerQA \citep{peerqa}, CovidQA \citep{covidqa}, and MLDR \citep{mldr} for document retrieval.
The document benchmarks cover financial reports, scientific papers, biomedical text, and long documents.
We use the candidate banks and relevance labels released by LMEB, with 5,873 queries in total \citep{lmeb}.
We further evaluate downstream question answering on LoCoMo and LongMemEval to assess whether retrieval improvements lead to better answers \citep{locomo,longmem}.

\noindent\textbf{Baselines.}
We use cosine similarity between the original embeddings as Base.
We compare NEAR with LC's local-affinity correction \citep{lc}, InvGC's reference-based embedding transformation \citep{invgc}, and DBSN's similarity normalization \citep{dbsn}.
Search-Adaptor provides a supervised comparison by learning embedding mappings from relevance labels \citep{searchadaptor}.
These same-backbone methods use the same frozen embeddings, data splits, and development-selection settings.
We also include BM25 \citep{bm25} as a lexical baseline, alongside ColBERTv2 \citep{colbertv2} and XTR \citep{xtr} as dedicated multi-vector retrievers.
The latter use their native encoders and retrieval configurations on the same candidate banks.

\noindent\textbf{Implementation Details.}
We evaluate nine frozen embedding models \citep{qwen,embeddinggemma,kalm,modernbert,nvretriever,nemotron8b}.
Following the LMEB evaluation settings, we report Recall@10 and NDCG@10, with Overall averaged equally across the eight tasks.
NEAR uses HNSW for retrieval \citep{hnsw}.
Residual depth, normalization, and fusion weight are selected using development NDCG@10.
Model configurations, fixed hyperparameters, and evaluation splits are detailed in Appendices~\ref{app:population} and \ref{app:selection}.

For downstream question answering, we compare Base with NEAR using Qwen3-30B-A3B \citep{qwen3} to generate answers and GPT-4o-mini \citep{gpt4omini} to judge correctness.
We report Accuracy and answer-text F1 at $K=10$.
The retrieval-budget analysis varies $K\in\{4,8,10,16\}$.

\suppressfloats[t]
\begin{table}[t]
\centering
\caption{\textbf{Retrieval quality across three Qwen scales on eight LMEB tasks (\%).} Cells report capped Recall@10 / NDCG@10. Overall reports the unweighted mean across the eight tasks.}
\label{tab:main}
\begingroup\nearUnifiedTableStyle\fontsize{8.0}{10}\selectfont\setlength{\tabcolsep}{1.8pt}
\newcommand{\nearMainHeader}[1]{%
  \begingroup\fontsize{7.2}{8.6}\selectfont\bfseries
  \settowidth{\dimen0}{#1}%
  \ifdim\dimen0>\linewidth
    \resizebox{\linewidth}{\height}{#1}%
  \else #1\fi\endgroup}
\begin{tabularx}{\linewidth}{@{\hspace{\nearTableEdgePadding}}>{\columncolor{white}[\nearTableEdgePadding][\tabcolsep]}l*{8}{>{\centering\arraybackslash}X}>{\columncolor{white}[\tabcolsep][\nearTableEdgePadding]\centering\arraybackslash}X@{\hspace{\nearTableEdgePadding}}}\specialrule{\heavyrulewidth}{0pt}{0pt}
\textbf{Method} & \nearMainHeader{LoCoMo} & \nearMainHeader{LongMemEval} & \nearMainHeader{REALTALK} & \nearMainHeader{ESG-Reports} & \nearMainHeader{QASPER} & \nearMainHeader{PeerQA} & \nearMainHeader{CovidQA} & \nearMainHeader{MLDR} & \nearMainHeader{Overall}\\\hline
\rowcolor{nearGroupTint}\multicolumn{10}{c}{\textbf{Standalone retrieval systems}}\\
BM25 & 34.7/25.3 & 73.4/62.7 & 27.7/21.7 & 49.8/34.3 & 55.1/32.2 & 26.7/20.3 & 89.8/73.6 & 99.0/83.4 & 57.0/44.2\\
ColBERTv2 & \mbox{63.0/50.0} & \mbox{87.5/80.0} & \mbox{54.3/45.8} & \mbox{56.9/45.0} & \mbox{75.5/52.0} & \mbox{45.3/36.3} & \mbox{95.2/84.4} & \mbox{96.0/85.4} & \mbox{71.7/59.9}\\
XTR & \mbox{43.5/31.6} & \mbox{89.4/79.1} & \mbox{42.7/33.1} & \mbox{62.9/51.9} & \mbox{71.2/48.8} & \mbox{38.8/29.8} & \mbox{93.9/83.3} & \mbox{98.0/88.2} & \mbox{67.5/55.7}\\
\addlinespace[3pt]
\rowcolor{nearGroupTint}\multicolumn{10}{c}{\textbf{Qwen3-Embedding-0.6B}}\\
Base & \mbox{46.4/33.9} & \mbox{84.8/72.8} & \mbox{46.4/37.4} & \mbox{\textbf{52.8}/40.1} & \mbox{69.7/47.0} & \mbox{36.2/26.7} & \mbox{91.4/73.1} & \mbox{\textbf{95.0}/76.4} & \mbox{65.3/50.9}\\
LC & \mbox{50.4/37.3} & \mbox{85.9/73.8} & \mbox{51.1/\textbf{41.0}} & \mbox{46.1/36.4} & \mbox{69.6/47.2} & \mbox{37.1/27.3} & \mbox{90.8/73.5} & \mbox{\textbf{95.0}/\textbf{76.9}} & \mbox{65.7/51.7}\\
InvGC & \mbox{48.1/35.0} & \mbox{\textbf{88.7}/77.0} & \mbox{49.3/39.7} & \mbox{45.6/36.2} & \mbox{69.8/47.1} & \mbox{36.7/27.4} & \mbox{92.2/74.5} & \mbox{\textbf{95.0}/76.4} & \mbox{65.7/51.7}\\
DBSN & \mbox{46.6/34.1} & \mbox{86.7/75.9} & \mbox{46.7/37.7} & \mbox{48.1/35.3} & \mbox{69.7/47.0} & \mbox{38.1/28.4} & \mbox{92.5/74.4} & \mbox{\textbf{95.0}/76.4} & \mbox{65.4/51.2}\\
Search-Adaptor & \mbox{51.9/38.0} & \mbox{88.4/\textbf{77.3}} & \mbox{49.3/39.3} & \mbox{41.4/30.8} & \mbox{\textbf{73.7}/\textbf{51.6}} & \mbox{38.6/28.2} & \mbox{\textbf{93.1}/\textbf{76.1}} & \mbox{\textbf{95.0}/76.7} & \mbox{66.4/52.2}\\
\rowcolor{nearTableTint}\textbf{NEAR} & \mbox{\textbf{55.7}/\textbf{41.5}} & \mbox{86.2/75.1} & \mbox{\textbf{51.2}/40.8} & \mbox{49.4/\textbf{40.5}} & \mbox{69.4/47.4} & \mbox{\textbf{39.8}/\textbf{29.0}} & \mbox{90.5/73.6} & \mbox{\textbf{95.0}/76.1} & \mbox{\textbf{67.2}/\textbf{53.0}}\\
\addlinespace[3pt]
\rowcolor{nearGroupTint}\multicolumn{10}{c}{\textbf{Qwen3-Embedding-4B}}\\
Base & \mbox{41.4/29.7} & \mbox{59.0/47.1} & \mbox{42.5/34.5} & \mbox{60.7/41.2} & \mbox{73.1/50.2} & \mbox{35.2/26.7} & \mbox{93.5/76.6} & \mbox{96.0/77.0} & \mbox{62.7/47.9}\\
LC & \mbox{44.6/32.2} & \mbox{59.0/47.1} & \mbox{44.1/34.7} & \mbox{56.2/41.5} & \mbox{73.0/50.1} & \mbox{37.4/28.0} & \mbox{93.1/76.7} & \mbox{96.0/77.7} & \mbox{62.9/48.5}\\
InvGC & \mbox{49.3/35.6} & \mbox{91.3/81.2} & \mbox{46.8/37.6} & \mbox{59.1/44.7} & \mbox{73.2/51.2} & \mbox{38.3/27.9} & \mbox{93.5/78.1} & \mbox{96.0/\textbf{78.5}} & \mbox{68.4/54.4}\\
DBSN & \mbox{51.7/38.1} & \mbox{91.7/80.5} & \mbox{\textbf{49.5}/\textbf{40.2}} & \mbox{58.4/42.6} & \mbox{73.1/50.2} & \mbox{\textbf{39.6}/29.1} & \mbox{93.4/76.7} & \mbox{96.0/77.0} & \mbox{69.2/54.3}\\
Search-Adaptor & \mbox{50.1/37.4} & \mbox{\textbf{93.6}/\textbf{83.1}} & \mbox{43.7/35.5} & \mbox{52.9/38.3} & \mbox{\textbf{75.7}/\textbf{54.8}} & \mbox{37.9/28.8} & \mbox{\textbf{94.0}/\textbf{79.1}} & \mbox{96.0/78.3} & \mbox{68.0/54.4}\\
\rowcolor{nearTableTint}\textbf{NEAR} & \mbox{\textbf{56.1}/\textbf{42.0}} & \mbox{89.7/78.5} & \mbox{47.2/37.7} & \mbox{\textbf{62.7}/\textbf{47.8}} & \mbox{72.8/50.9} & \mbox{38.9/\textbf{30.3}} & \mbox{93.8/77.6} & \mbox{\textbf{97.0}/77.8} & \mbox{\textbf{69.8}/\textbf{55.3}}\\
\addlinespace[3pt]
\rowcolor{nearGroupTint}\multicolumn{10}{c}{\textbf{Qwen3-Embedding-8B}}\\
Base & \mbox{49.2/36.2} & \mbox{87.9/77.4} & \mbox{49.8/40.7} & \mbox{62.2/43.0} & \mbox{70.4/47.4} & \mbox{40.8/30.1} & \mbox{94.0/79.5} & \mbox{95.0/77.4} & \mbox{68.6/53.9}\\
LC & \mbox{50.7/37.1} & \mbox{87.9/77.3} & \mbox{50.5/40.2} & \mbox{63.1/44.2} & \mbox{70.3/47.3} & \mbox{42.1/30.2} & \mbox{94.0/79.3} & \mbox{\textbf{96.0}/78.1} & \mbox{69.3/54.2}\\
InvGC & \mbox{52.0/38.8} & \mbox{\textbf{94.7}/84.2} & \mbox{51.9/\textbf{42.4}} & \mbox{56.2/37.9} & \mbox{71.6/48.3} & \mbox{38.2/28.6} & \mbox{94.1/\textbf{80.0}} & \mbox{\textbf{96.0}/77.0} & \mbox{69.4/54.6}\\
DBSN & \mbox{51.7/39.2} & \mbox{\textbf{94.7}/\textbf{84.7}} & \mbox{51.3/42.3} & \mbox{62.3/42.4} & \mbox{70.4/47.4} & \mbox{41.4/30.7} & \mbox{93.7/78.9} & \mbox{95.0/77.4} & \mbox{70.1/55.4}\\
Search-Adaptor & \mbox{54.7/41.3} & \mbox{93.1/83.0} & \mbox{50.6/41.0} & \mbox{52.9/37.2} & \mbox{\textbf{74.9}/\textbf{52.1}} & \mbox{41.2/30.8} & \mbox{\textbf{94.6}/79.9} & \mbox{\textbf{96.0}/\textbf{78.7}} & \mbox{69.7/55.5}\\
\rowcolor{nearTableTint}\textbf{NEAR} & \mbox{\textbf{59.9}/\textbf{45.3}} & \mbox{90.7/81.6} & \mbox{\textbf{52.6}/42.3} & \mbox{\textbf{68.0}/\textbf{49.9}} & \mbox{69.8/47.1} & \mbox{\textbf{43.1}/\textbf{32.1}} & \mbox{93.9/79.1} & \mbox{\textbf{96.0}/78.5} & \mbox{\textbf{71.8}/\textbf{57.0}}\\
\specialrule{\heavyrulewidth}{0pt}{0pt}\end{tabularx}\endgroup

\end{table}

\subsection{Main Retrieval Results}
\noindent\textbf{Overall retrieval performance.}
Table~\ref{tab:main} shows that NEAR achieves the highest average Recall@10 and NDCG@10 among the evaluated same-backbone methods at all three Qwen3-Embedding sizes.
For Qwen-4B, average Recall@10 increases from 62.67 to 69.76, while NDCG@10 rises from 47.88 to 55.33.
On Qwen-8B, NEAR exceeds the strongest same-backbone baseline for each metric by 1.68 percentage points in average Recall@10 and 1.50 points in average NDCG@10.

\noindent\textbf{Performance across tasks.}
On LoCoMo, NEAR achieves the best Recall@10 and NDCG@10 among same-backbone methods at all three Qwen sizes.
On LongMemEval, it also improves both metrics over Base.
For Qwen-4B, the two metrics increase by 14.76 and 12.34 percentage points on LoCoMo and by 30.70 and 31.39 points on LongMemEval, respectively.
On LongMemEval, Qwen-8B with NEAR also outperforms ColBERTv2 and XTR on both metrics.
For document retrieval, NEAR achieves the best same-backbone scores on both metrics for ESG-Reports with Qwen-4B and Qwen-8B, and the highest same-backbone NDCG@10 for PeerQA at all three Qwen sizes.

\noindent\textbf{Results across embedding models.}
NEAR's improvements are not limited to the Qwen series.
Table~\ref{tab:crossmodel} shows that NEAR improves average Recall@10 and NDCG@10 over Base for all nine frozen embedding models, including both small and large models.
For example, the two metrics increase by 3.43 and 1.95 percentage points on EmbeddingGemma-300M and by 2.52 and 2.11 points on KaLM-Embedding-Gemma3-12B, respectively.
Complete comparisons with the other methods for each model appear in Appendix~\ref{app:allencoders}.

\begin{table}[tb]
\centering
\caption{\textbf{Cross-model gains.} Eight-task mean capped Recall@10 and NDCG@10 (\%).}
\label{tab:crossmodel}
\begingroup\nearUnifiedTableStyle\centering\fontsize{8.2}{9.8}\selectfont\setlength{\tabcolsep}{2.5pt}
\begin{tabularx}{\linewidth}{@{\hspace{\nearTableEdgePadding}}p{.345\linewidth}>{\centering\arraybackslash}X>{\columncolor{nearTableTint}\centering\arraybackslash}X>{\centering\arraybackslash}X>{\columncolor{nearTableTint}[\tabcolsep][\nearTableEdgePadding]\centering\arraybackslash}X@{\hspace{\nearTableEdgePadding}}}\toprule
& \multicolumn{2}{c}{\textbf{Recall@10}} & \multicolumn{2}{c}{\textbf{NDCG@10}}\\
\cmidrule(l){2-3}\cmidrule(l){4-5}
\textbf{Model} & Base & \textbf{NEAR} & Base & \textbf{NEAR}\\\midrule
Qwen3-Embedding-0.6B & 65.34 & \textbf{67.17} & 50.93 & \textbf{52.99}\\
Qwen3-Embedding-4B & 62.67 & \textbf{69.76} & 47.88 & \textbf{55.33}\\
Qwen3-Embedding-8B & 68.65 & \textbf{71.76} & 53.95 & \textbf{57.00}\\
Llama-Nemotron-Embed-1B-v2 & 74.61 & \textbf{75.05} & 61.25 & \textbf{61.55}\\
Llama-NV-Embed-Reasoning-3B & 77.50 & \textbf{78.78} & 60.96 & \textbf{62.40}\\
Llama-Embed-Nemotron-8B & 76.51 & \textbf{77.88} & 62.73 & \textbf{63.71}\\
EmbeddingGemma-300M & 72.66 & \textbf{76.09} & 59.36 & \textbf{61.31}\\
KaLM-Embedding-Gemma3-12B & 72.67 & \textbf{75.19} & 57.11 & \textbf{59.21}\\
ModernBERT-Embed-Base & 69.76 & \textbf{70.88} & 55.72 & \textbf{57.08}\\
\bottomrule\end{tabularx}\par\endgroup

\end{table}


\subsection{Downstream Question Answering}
We further examine whether retrieval improvements translate into better downstream answers.
Table~\ref{tab:native-qa} compares Base and NEAR across nine backbones using the original conversation histories of LoCoMo and LongMemEval.
All comparisons use the same answering model and prompts, with a fixed budget of ten retrieved memories.

\begin{table}[!htbp]
\centering
\caption{\textbf{Downstream QA at Top-10 (\%).} NEAR scores are followed by blue parenthesized changes (NEAR $-$ Base, in percentage points). Category columns report accuracy.}
\label{tab:native-qa}
\definecolor{deltaInk}{RGB}{55,95,143}

\newcommand{\nearQACell}[4]{%
  \begingroup%
  \settowidth{\dimen0}{#1}%
  \makebox[\dimen0][r]{#3}\hspace{0.6pt}%
  {\fontsize{6.4}{7}\selectfont%
   \settowidth{\dimen0}{#2}%
   \textcolor{deltaInk}{\makebox[\dimen0][l]{#4}}}%
  \endgroup}

\begingroup\nearUnifiedTableStyle\fontsize{8.0}{10}\selectfont
\setlength{\tabcolsep}{0.5pt}\renewcommand{\arraystretch}{1.03}
\begin{tabular*}{\linewidth}{@{\hspace{\nearTableEdgePadding}\extracolsep{\fill}}l*{6}{c}@{\hspace{\nearTableEdgePadding}}}
\toprule
\multicolumn{5}{@{\hspace{\nearTableEdgePadding}}l}{\textbf{(a) LoCoMo}} & \multicolumn{2}{c}{\hspace*{-0.977165pt}\textbf{Overall}\hspace*{0.977165pt}}\\
\cmidrule(l{8.447413pt}r{10.393412pt}){6-7}
\textbf{Model} & S-hop & Multi & Temp. & Open & \textbf{Acc.} & \textbf{F1}\\\midrule
Qwen3-Embedding-0.6B & \nearQACell{99.9}{(+18.1)}{83.8}{(+7.4)} & \nearQACell{99.9}{(+14.2)}{64.9}{(+6.0)} & \nearQACell{99.9}{(+4.4)}{49.2}{(+4.4)} & \nearQACell{99.9}{(+12.5)}{57.3}{(+6.2)} & \nearQACell{99.9}{(+13.3)}{71.5}{(+6.4)} & \nearQACell{99.9}{(+3.7)}{34.6}{(+3.7)}\\
Qwen3-Embedding-4B & \nearQACell{99.9}{(+18.1)}{85.7}{(+9.2)} & \nearQACell{99.9}{(+14.2)}{66.3}{(+14.2)} & \nearQACell{99.9}{(+4.4)}{46.7}{(+8.1)} & \nearQACell{99.9}{(+12.5)}{58.3}{(+5.2)} & \nearQACell{99.9}{(+13.3)}{72.3}{(+9.6)} & \nearQACell{99.9}{(+3.7)}{35.5}{(+4.6)}\\
Qwen3-Embedding-8B & \nearQACell{99.9}{(+18.1)}{85.1}{(+1.7)} & \nearQACell{99.9}{(+14.2)}{63.5}{(+1.1)} & \nearQACell{99.9}{(+4.4)}{46.7}{(+4.0)} & \nearQACell{99.9}{(+12.5)}{60.4}{(+12.5)} & \nearQACell{99.9}{(+13.3)}{71.6}{(+2.7)} & \nearQACell{99.9}{(+3.7)}{34.2}{(+0.8)}\\
\addlinespace[1.5pt]
Llama-Nemotron-Embed-1B-v2 & \nearQACell{99.9}{(+18.1)}{85.1}{(+5.8)} & \nearQACell{99.9}{(+14.2)}{55.0}{(+3.5)} & \nearQACell{99.9}{(+4.4)}{42.1}{(+0.3)} & \nearQACell{99.9}{(+12.5)}{57.3}{(+7.3)} & \nearQACell{99.9}{(+13.3)}{68.9}{(+4.4)} & \nearQACell{99.9}{(+3.7)}{34.0}{(+2.3)}\\
Llama-NV-Embed-Reasoning-3B & \nearQACell{99.9}{(+18.1)}{54.3}{(+18.1)} & \nearQACell{99.9}{(+14.2)}{35.5}{(+9.9)} & \nearQACell{99.9}{(+4.4)}{21.8}{(+6.2)} & \nearQACell{99.9}{(+12.5)}{40.6}{(+5.2)} & \nearQACell{99.9}{(+13.3)}{43.2}{(+13.3)} & \nearQACell{99.9}{(+3.7)}{21.5}{(+7.3)}\\
Llama-Embed-Nemotron-8B & \nearQACell{99.9}{(+18.1)}{89.2}{(+2.5)} & \nearQACell{99.9}{(+14.2)}{64.9}{(+6.0)} & \nearQACell{99.9}{(+4.4)}{47.7}{(+2.8)} & \nearQACell{99.9}{(+12.5)}{52.1}{(\textminus{}1.0)} & \nearQACell{99.9}{(+13.3)}{73.8}{(+3.0)} & \nearQACell{99.9}{(+3.7)}{36.7}{(+1.9)}\\
\addlinespace[1.5pt]
EmbeddingGemma-300M & \nearQACell{99.9}{(+18.1)}{88.0}{(+4.0)} & \nearQACell{99.9}{(+14.2)}{62.4}{(+6.4)} & \nearQACell{99.9}{(+4.4)}{47.4}{(+3.4)} & \nearQACell{99.9}{(+12.5)}{55.2}{(0.0)} & \nearQACell{99.9}{(+13.3)}{72.8}{(+4.1)} & \nearQACell{99.9}{(+3.7)}{35.9}{(+2.2)}\\
KaLM-Embedding-Gemma3-12B & \nearQACell{99.9}{(+18.1)}{88.6}{(\textminus{}0.1)} & \nearQACell{99.9}{(+14.2)}{65.6}{(+2.8)} & \nearQACell{99.9}{(+4.4)}{48.3}{(+1.6)} & \nearQACell{99.9}{(+12.5)}{53.1}{(\textminus{}1.0)} & \nearQACell{99.9}{(+13.3)}{73.8}{(+0.7)} & \nearQACell{99.9}{(+3.7)}{36.5}{(+0.1)}\\
ModernBERT-Embed-Base & \nearQACell{99.9}{(+18.1)}{70.5}{(+18.8)} & \nearQACell{99.9}{(+14.2)}{47.5}{(+11.0)} & \nearQACell{99.9}{(+4.4)}{36.1}{(+8.4)} & \nearQACell{99.9}{(+12.5)}{51.0}{(+12.5)} & \nearQACell{99.9}{(+13.3)}{57.9}{(+14.8)} & \nearQACell{99.9}{(+3.7)}{29.8}{(+7.9)}\\
\midrule\end{tabular*}\endgroup
\par\vspace{-2pt}

\begingroup\nearUnifiedTableStyle\fontsize{8.0}{10}\selectfont
\setlength{\tabcolsep}{0.5pt}\renewcommand{\arraystretch}{1.03}
\begin{tabular*}{\linewidth}{@{\hspace{\nearTableEdgePadding}\extracolsep{\fill}}l*{8}{c}@{\hspace{\nearTableEdgePadding}}}

\multicolumn{7}{@{\hspace{\nearTableEdgePadding}}l}{\textbf{(b) LongMemEval}} & \multicolumn{2}{c}{\hspace*{-1.780201pt}\textbf{Overall}\hspace*{1.780201pt}}\\
\cmidrule(l{0.691989pt}r{4.244090pt}){8-9}
\textbf{Model} & User & Asst. & Pref. & Multi & Temp. & Upd. & \textbf{Acc.} & \textbf{F1}\\\midrule
Qwen3-Embedding-0.6B & \nearQACell{99.9}{(\textminus{}1.4)}{91.4}{(+5.7)} & \nearQACell{100.0}{(\textminus{}1.8)}{98.2}{(+1.8)} & \nearQACell{99.9}{(\textminus{}10.0)}{70.0}{(+10.0)} & \nearQACell{99.9}{(\textminus{}0.8)}{47.4}{(\textminus{}0.8)} & \nearQACell{99.9}{(\textminus{}0.8)}{54.1}{(+4.5)} & \nearQACell{99.9}{(\textminus{}3.8)}{80.8}{(\textminus{}3.8)} & \nearQACell{99.9}{(\textminus{}0.8)}{67.6}{(+2.0)} & \nearQACell{99.9}{(\textminus{}0.2)}{11.5}{(+0.2)}\\
Qwen3-Embedding-4B & \nearQACell{99.9}{(\textminus{}1.4)}{97.1}{(+2.9)} & \nearQACell{100.0}{(\textminus{}1.8)}{98.2}{(0.0)} & \nearQACell{99.9}{(\textminus{}10.0)}{70.0}{(+16.7)} & \nearQACell{99.9}{(\textminus{}0.8)}{52.6}{(+2.3)} & \nearQACell{99.9}{(\textminus{}0.8)}{60.2}{(\textminus{}0.8)} & \nearQACell{99.9}{(\textminus{}3.8)}{85.9}{(+6.4)} & \nearQACell{99.9}{(\textminus{}0.8)}{72.1}{(+2.8)} & \nearQACell{99.9}{(\textminus{}0.2)}{12.3}{(+0.1)}\\
Qwen3-Embedding-8B & \nearQACell{99.9}{(\textminus{}1.4)}{95.7}{(0.0)} & \nearQACell{100.0}{(\textminus{}1.8)}{98.2}{(+1.8)} & \nearQACell{99.9}{(\textminus{}10.0)}{76.7}{(+6.7)} & \nearQACell{99.9}{(\textminus{}0.8)}{60.9}{(0.0)} & \nearQACell{99.9}{(\textminus{}0.8)}{58.6}{(\textminus{}1.5)} & \nearQACell{99.9}{(\textminus{}3.8)}{84.6}{(+3.8)} & \nearQACell{99.9}{(\textminus{}0.8)}{74.0}{(+0.8)} & \nearQACell{99.9}{(\textminus{}0.2)}{11.6}{(\textminus{}0.2)}\\
\addlinespace[1.5pt]
Llama-Nemotron-Embed-1B-v2 & \nearQACell{99.9}{(\textminus{}1.4)}{88.6}{(+4.3)} & \nearQACell{100.0}{(\textminus{}1.8)}{96.4}{(0.0)} & \nearQACell{99.9}{(\textminus{}10.0)}{63.3}{(+6.7)} & \nearQACell{99.9}{(\textminus{}0.8)}{42.9}{(+3.0)} & \nearQACell{99.9}{(\textminus{}0.8)}{49.6}{(+6.0)} & \nearQACell{99.9}{(\textminus{}3.8)}{85.9}{(+1.3)} & \nearQACell{99.9}{(\textminus{}0.8)}{65.0}{(+3.6)} & \nearQACell{99.9}{(\textminus{}0.2)}{11.8}{(+0.2)}\\
Llama-NV-Embed-Reasoning-3B & \nearQACell{99.9}{(\textminus{}1.4)}{82.9}{(+1.4)} & \nearQACell{100.0}{(\textminus{}1.8)}{100.0}{(+1.8)} & \nearQACell{99.9}{(\textminus{}10.0)}{66.7}{(\textminus{}10.0)} & \nearQACell{99.9}{(\textminus{}0.8)}{36.1}{(\textminus{}0.8)} & \nearQACell{99.9}{(\textminus{}0.8)}{44.4}{(\textminus{}3.0)} & \nearQACell{99.9}{(\textminus{}3.8)}{79.5}{(+2.6)} & \nearQACell{99.9}{(\textminus{}0.8)}{60.6}{(\textminus{}0.8)} & \nearQACell{99.9}{(\textminus{}0.2)}{10.9}{(+0.4)}\\
Llama-Embed-Nemotron-8B & \nearQACell{99.9}{(\textminus{}1.4)}{95.7}{(0.0)} & \nearQACell{100.0}{(\textminus{}1.8)}{98.2}{(\textminus{}1.8)} & \nearQACell{99.9}{(\textminus{}10.0)}{70.0}{(\textminus{}10.0)} & \nearQACell{99.9}{(\textminus{}0.8)}{51.9}{(+3.0)} & \nearQACell{99.9}{(\textminus{}0.8)}{55.6}{(+2.3)} & \nearQACell{99.9}{(\textminus{}3.8)}{85.9}{(+3.8)} & \nearQACell{99.9}{(\textminus{}0.8)}{70.6}{(+1.2)} & \nearQACell{99.9}{(\textminus{}0.2)}{12.0}{(0.0)}\\
\addlinespace[1.5pt]
EmbeddingGemma-300M & \nearQACell{99.9}{(\textminus{}1.4)}{95.7}{(+2.9)} & \nearQACell{100.0}{(\textminus{}1.8)}{98.2}{(+1.8)} & \nearQACell{99.9}{(\textminus{}10.0)}{76.7}{(+3.3)} & \nearQACell{99.9}{(\textminus{}0.8)}{48.1}{(\textminus{}2.3)} & \nearQACell{99.9}{(\textminus{}0.8)}{51.9}{(\textminus{}6.8)} & \nearQACell{99.9}{(\textminus{}3.8)}{85.9}{(+6.4)} & \nearQACell{99.9}{(\textminus{}0.8)}{69.0}{(\textminus{}0.6)} & \nearQACell{99.9}{(\textminus{}0.2)}{11.7}{(+0.3)}\\
KaLM-Embedding-Gemma3-12B & \nearQACell{99.9}{(\textminus{}1.4)}{95.7}{(\textminus{}1.4)} & \nearQACell{100.0}{(\textminus{}1.8)}{100.0}{(+1.8)} & \nearQACell{99.9}{(\textminus{}10.0)}{76.7}{(+6.7)} & \nearQACell{99.9}{(\textminus{}0.8)}{66.2}{(+5.3)} & \nearQACell{99.9}{(\textminus{}0.8)}{66.2}{(0.0)} & \nearQACell{99.9}{(\textminus{}3.8)}{83.3}{(\textminus{}1.3)} & \nearQACell{99.9}{(\textminus{}0.8)}{77.4}{(+1.6)} & \nearQACell{99.9}{(\textminus{}0.2)}{12.7}{(+0.3)}\\
ModernBERT-Embed-Base & \nearQACell{99.9}{(\textminus{}1.4)}{90.0}{(+1.4)} & \nearQACell{100.0}{(\textminus{}1.8)}{94.6}{(\textminus{}3.6)} & \nearQACell{99.9}{(\textminus{}10.0)}{63.3}{(0.0)} & \nearQACell{99.9}{(\textminus{}0.8)}{46.6}{(+0.8)} & \nearQACell{99.9}{(\textminus{}0.8)}{52.6}{(+3.8)} & \nearQACell{99.9}{(\textminus{}3.8)}{76.9}{(+1.3)} & \nearQACell{99.9}{(\textminus{}0.8)}{65.4}{(+1.2)} & \nearQACell{99.9}{(\textminus{}0.2)}{11.4}{(+0.2)}\\
\bottomrule\end{tabular*}\endgroup

\end{table}

On LoCoMo, NEAR improves both Accuracy and F1 across all nine backbones.
For Qwen-4B, Accuracy increases from 62.7 to 72.3, while F1 rises from 30.9 to 35.5.
The improvements extend across model families: Llama-3B gains 13.3 and 7.3 percentage points on the two metrics, respectively.
These results show that NEAR's retrieval gains can translate into better answers to questions about conversation histories.

On LongMemEval, NEAR improves Accuracy on seven of the nine backbones, including Qwen-4B from 69.3 to 72.1 and KaLM-12B from 75.8 to 77.4; Llama-3B and EmbeddingGemma show small decreases.
Supplementary lexical F1 does not always follow Accuracy: Qwen-8B gains 0.8 percentage points in Accuracy while losing 0.2 points in F1, reflecting the distinction between answer correctness and lexical overlap.

\subsection{Ablation Study}
\label{sec:neighborhood-references}
We conduct ablations on all eight retrieval tasks using the three Qwen3-Embedding sizes and Llama-3B.
Each variant retains the depth, residual normalization, and fusion weight selected for full NEAR on development data, without retuning, and preserves original embedding matching.
Table~\ref{tab:selectedmechanism} summarizes the results, and Appendix~\ref{app:mechanisms} details the controls.

\begin{table}[!ht]
\centering
\caption{\textbf{Effect of neighborhood references (\%).} Eight-task mean capped Recall@10 (R@10) and NDCG@10 (N@10). Bold marks column bests.}
\label{tab:selectedmechanism}
\begingroup\nearUnifiedTableStyle\fontsize{8.2}{9.8}\selectfont\setlength{\tabcolsep}{1pt}
\begin{tabularx}{\linewidth}{@{\hspace{\nearTableEdgePadding}}>{\columncolor{white}[\nearTableEdgePadding][\tabcolsep]}p{70pt}*{6}{>{\centering\arraybackslash}X}>{\centering\arraybackslash}p{48pt}>{\columncolor{white}[\tabcolsep][\nearTableEdgePadding]\centering\arraybackslash}p{48pt}@{\hspace{\nearTableEdgePadding}}}\toprule
& \multicolumn{2}{c}{{\fontsize{6.8}{8.2}\selectfont\textbf{Qwen3-Embedding-0.6B}}} & \multicolumn{2}{c}{{\fontsize{6.8}{8.2}\selectfont\textbf{Qwen3-Embedding-4B}}} & \multicolumn{2}{c}{{\fontsize{6.8}{8.2}\selectfont\textbf{Qwen3-Embedding-8B}}} & \multicolumn{2}{c}{{\fontsize{6.8}{8.2}\selectfont\textbf{Llama-NV-Embed-Reasoning-3B}}}\\
\cmidrule(lr){2-3}\cmidrule(lr){4-5}\cmidrule(lr){6-7}\cmidrule(l){8-9}
\textbf{Variant} & R@10 & N@10 & R@10 & N@10 & R@10 & N@10 & R@10 & N@10\\
\midrule
Global reference & 66.18 & 51.84 & 65.12 & 51.14 & 69.69 & 54.93 & 77.85 & 61.23\\
Shuffled references & 65.80 & 51.80 & 68.14 & 53.96 & 70.22 & 55.72 & 77.69 & 61.79\\
Memory only & 65.97 & 52.12 & 67.54 & 53.76 & 70.86 & 56.18 & 78.69 & 62.13\\
Query only & 66.74 & 52.24 & 67.94 & 53.17 & 70.19 & 55.15 & 77.77 & 60.80\\
\rowcolor{nearTableTint}\textbf{NEAR} & \textbf{67.17} & \textbf{52.99} & \textbf{69.76} & \textbf{55.33} & \textbf{71.76} & \textbf{57.00} & \textbf{78.78} & \textbf{62.40}\\
\bottomrule\end{tabularx}\par\endgroup

\end{table}

\begin{samepage}
\noindent\textbf{Local references and their correspondence.}
We first replace per-memory local references with the memory bank's global mean direction.
Full NEAR outperforms this variant on both metrics for all four backbones, with average gains of $2.16$ and $2.15$ percentage points in Recall@10 and NDCG@10, respectively.
Keeping the local references but shuffling their assignments to memories reduces the two metrics by $1.40$ and $1.12$ points on average.
These results indicate that the gains depend not just on introducing reference directions, but also on their locality and correspondence with individual memories.
\par\end{samepage}

\noindent\textbf{Query- and memory-side references.}
We construct residuals on only the memory or query side, retaining the original embeddings on the other side.
Full NEAR outperforms both one-sided variants on all four backbones: average Recall@10 and NDCG@10 increase by $1.10$ and $0.88$ points over memory-only matching, and by $1.21$ and $1.59$ points over query-only matching.
These results support using both query- and memory-side references to construct complementary matching representations.

\noindent\textbf{Case study.}
In the LoCoMo example in Figure~\ref{fig:case-study}, Base retrieves personal updates and endorsement experiences but misses Tim's advice to align endorsements with John's values and brand. Local references define the projection directions. Here, distractors' raw scores fall more than the correct evidence's after projection, indicating that the removed components contribute to their original ranking advantage. The evidence outscores Base's top ten in the first residual layer. Fusing all residual scores with original matching moves it from rank 75 to rank 1.

\FloatBarrier

\begingroup
\setlength{\intextsep}{6pt plus 2pt minus 2pt}
\captionsetup[figure]{skip=6pt}
\Needspace{27\baselineskip}
\subsection{Sensitivity Analysis}
\label{sec:parameter-sensitivity}

\noindent\textbf{Fusion weight and residual construction.}
We study the effects of fusion weight, residual normalization, and depth using Qwen3-Embedding-4B on the complete query sets of LoCoMo and LongMemEval.
We vary the total residual weight over $\alpha\in\{0.25,0.5,1,2,4\}$.
Different depths use the same total residual weight, with all other settings held fixed.

\begin{figure}[H]
\centering
\includegraphics[width=\linewidth]{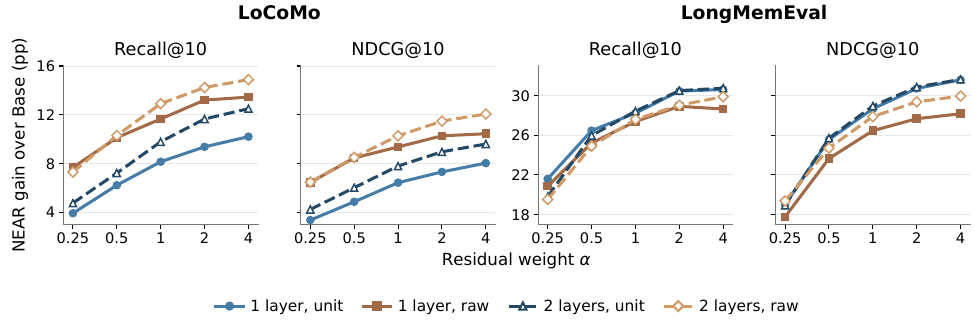}
\caption{\textbf{Sensitivity to residual weight and construction.} Each curve reports the NEAR gain over Base for Qwen3-Embedding-4B. Unit and raw denote normalized and unnormalized scoring views; layer choices use the same total residual weight $\alpha$. Vertical ranges differ by task.}
\label{fig:parameter-sensitivity}
\end{figure}

As shown in Figure~\ref{fig:parameter-sensitivity}, all displayed configurations outperform Base in both Recall@10 and NDCG@10 on both tasks throughout the tested weight range.
Performance generally improves as the residual weight increases and levels off at larger weights.
The tasks favor different residual constructions: unnormalized residuals perform better on LoCoMo, whereas normalized residuals achieve higher NDCG@10 at moderate and large weights on LongMemEval.
Thus, NEAR maintains gains over Base across multiple residual constructions and fusion weights, rather than only at a particular configuration.

\begin{figure}[!htbp]
\centering
\includegraphics[width=\linewidth]{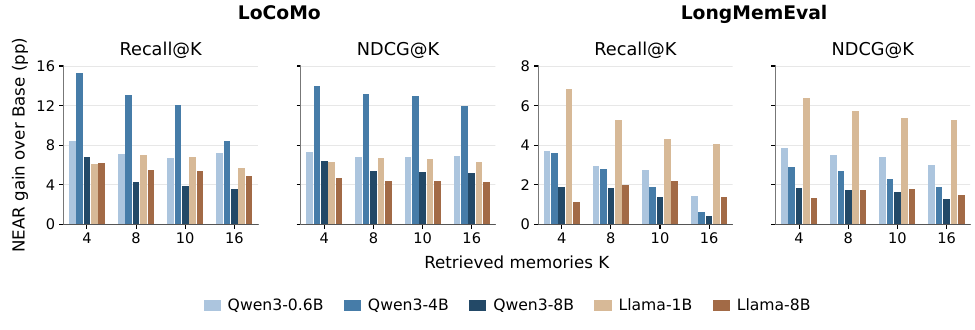}
\caption{\textbf{Gains across retrieved-memory budgets.} NEAR minus Base across five embedding models; vertical ranges differ by task.}
\label{fig:native-topk}
\end{figure}

\noindent\textbf{Retrieved memory budget.}
We further vary the number of returned memories over $K\in\{4,8,10,16\}$ using the original conversational candidate banks of LoCoMo and LongMemEval, with the remaining retrieval configuration held fixed.
Across the five embedding models shown in Figure~\ref{fig:native-topk}, NEAR improves Recall@$K$ and NDCG@$K$ over Base at every tested budget on both datasets.
The retrieval gains therefore extend beyond the Top-10 setting used in the main comparison.
These curves measure evidence retrieval; the Top-10 QA results in Table~\ref{tab:native-qa} separately assess downstream answer quality.
\FloatBarrier
\endgroup

\subsection{Retrieval Cost Analysis}
We compare retrieval quality, storage, and query latency using Qwen-0.6B/4B on LoCoMo and LongMemEval.
Quality is evaluated on the complete tasks, while storage and latency are measured on four representative candidate banks.
To examine the effect of index quantization, we compare Base and NEAR in FP32 and with four-bit scalar quantization (SQ4).
All indexes use the same HNSW settings \citep{hnsw,faiss}.
Query latency includes vector preparation and ANN search, but excludes encoding. Appendix~\ref{app:cost} provides additional details.

\begin{table}[!ht]
\centering
\captionsetup{width=\linewidth}
\caption{\textbf{Effect of index quantization.} Unmarked rows use FP32. Storage is relative to FP32 Base, and query latency excludes encoding.}
\label{tab:costselected}
\begingroup\nearUnifiedTableStyle\fontsize{8.2}{9.8}\selectfont\renewcommand{\arraystretch}{1.13}\setlength{\tabcolsep}{2pt}
\noindent\makebox[\linewidth][c]{\begin{tabularx}{\linewidth}{@{\hspace{\nearTableEdgePadding}}>{\columncolor{white}[\nearTableEdgePadding][\tabcolsep]}l*{3}{>{\centering\arraybackslash}X}>{\columncolor{white}[\tabcolsep][\nearTableEdgePadding]\centering\arraybackslash}X@{\hspace{\nearTableEdgePadding}}}\toprule
\textbf{Method} & Recall@10 & NDCG@10 & Storage ($\times$Base) & Latency (ms)\\
\midrule
Base & 57.91 & 45.88 & 1.00 & 0.242\\
Base (SQ4) & 57.81 & 45.82 & 0.17 & 0.239\\
\addlinespace[2pt]
\rowcolor{nearTableTint}\textbf{NEAR} & 71.82 & 59.21 & 2.02 & 0.579\\
\rowcolor{nearTableTint}\textbf{NEAR (SQ4)} & 71.97 & 59.16 & 0.36 & 0.590\\
\bottomrule\end{tabularx}}\par\endgroup

\end{table}

\noindent\textbf{Retrieval quality and latency.}
Table~\ref{tab:costselected} shows that uncompressed NEAR reaches mean Recall@10 and NDCG@10 of $71.82$ and $59.21$, compared with $57.91$ and $45.88$ for Base.
Median query latency increases from $0.242$ms to $0.579$ms, an additional $0.337$ms, or approximately $2.39$ times the Base latency.
NEAR therefore trades additional retrieval computation for improved quality rather than reducing query latency.

\noindent\textbf{Offline construction.}
On the same representative banks of 476 or 629 memories, building NEAR's references, scales, keys, and FP32 index takes $0.129$--$1.147$ seconds, compared with $0.040$--$0.140$ seconds for Base. NEAR with SQ4 takes $0.244$--$1.988$ seconds, including quantizer training. These are per-bank build times from cached embeddings, separate from query latency; the index and reference table are reused across queries. Appendix~\ref{app:cost} reports detailed construction times for all compared methods.

\noindent\textbf{Storage and quantization.}
Uncompressed NEAR has a median total-storage ratio of $2.02$ relative to FP32 Base.
SQ4 reduces NEAR's total assets by a median $82.7\%$, while mean Recall@10 changes from $71.82$ to $71.97$ and NDCG@10 from $59.21$ to $59.16$.
Its total assets are a median $36.0\%$ of FP32 Base; quantizing Base itself reduces its assets to $16.7\%$, with mean NDCG@10 of $45.82$.
Quantization thus substantially reduces NEAR's storage requirements while retaining its retrieval quality, although its assets remain larger than those of quantized Base.
NEAR's median query latency with SQ4 is $0.590$ms, close to its uncompressed latency, so the measured benefit is primarily in storage.
These timings characterize cached-vector retrieval on a shared CPU host, not end-to-end question answering.

\section{Conclusion}
\looseness=-1
We studied the difficulty of retrieving correct evidence from memory records with shared topics, entities, or events.
Such shared contents would make records without the required evidence more likely to outrank the correct evidence when ranking relied only on query--memory similarity.
Our key observation was that these similar records were not only competing candidates; they also provided a local neighborhood that could serve as a reference for each memory.
In this paper, we proposed NEAR, a training-free framework that turned this potential source of confusion into useful matching information by constructing memory-specific local references.
NEAR obtained query references from a precomputed table and combined residual representations with the original embeddings in a fixed index for a single approximate nearest-neighbor search.
Experiments on eight long-context retrieval benchmarks showed improvements in mean Recall@10 and NDCG@10 across nine embedding models, with further gains in downstream question answering.
In future work, we will extend NEAR to frequently updated memory banks and explore more compact reference representations.

\par\endgroup

\clearpage
\begingroup\raggedright
\bibliography{references}

\begin{thebibliography}{41}
\providecommand{\natexlab}[1]{#1}
\providecommand{\url}[1]{\texttt{#1}}
\expandafter\ifx\csname urlstyle\endcsname\relax
  \providecommand{\doi}[1]{doi: #1}\else
  \providecommand{\doi}{doi: \begingroup \urlstyle{rm}\Url}\fi

\bibitem[Ashok et~al.(2026)Ashok, Nair, Al-Darabsah, Teo, Agarwal, and
  May]{ashok2026sharpening}
Dhananjay Ashok, Suraj Nair, Mutasem Al-Darabsah, Choon~Hui Teo, Tarun Agarwal,
  and Jonathan May.
\newblock A representation sharpening framework for zero shot dense retrieval.
\newblock In \emph{Proceedings of the 19th Conference of the European Chapter
  of the Association for Computational Linguistics (Volume 1: Long Papers)},
  2026.

\bibitem[Babakhin et~al.(2025)Babakhin, Osmulski, Ak, Moreira, Xu, Schifferer,
  Liu, and Oldridge]{nemotron8b}
Yauhen Babakhin, Radek Osmulski, Ronay Ak, Gabriel Moreira, Mengyao Xu,
  Benedikt Schifferer, Bo~Liu, and Even Oldridge.
\newblock Llama-embed-nemotron-8b: A universal text embedding model for
  multilingual and cross-lingual tasks.
\newblock \emph{arXiv preprint arXiv:2511.07025}, 2025.

\bibitem[Baumg{\"a}rtner et~al.(2025)Baumg{\"a}rtner, Briscoe, and
  Gurevych]{peerqa}
Tim Baumg{\"a}rtner, Ted Briscoe, and Iryna Gurevych.
\newblock {PeerQA}: A scientific question answering dataset from peer reviews.
\newblock In \emph{Proceedings of the 2025 Conference of the Nations of the
  Americas Chapter of the Association for Computational Linguistics: Human
  Language Technologies (Volume 1: Long Papers)}, 2025.

\bibitem[Chen et~al.(2024)Chen, Xiao, Zhang, Luo, Lian, and Liu]{mldr}
Jianlyu Chen, Shitao Xiao, Peitian Zhang, Kun Luo, Defu Lian, and Zheng Liu.
\newblock {M3-Embedding}: Multi-linguality, multi-functionality,
  multi-granularity text embeddings through self-knowledge distillation.
\newblock In \emph{Findings of the Association for Computational Linguistics:
  ACL 2024}, pp.\  2318--2335, 2024.

\bibitem[Chowdhury et~al.(2024)Chowdhury, Wang, Shenoy, Kiela, Schwettmann, and
  Thrush]{nnn2024}
Neil Chowdhury, Franklin Wang, Sumedh Shenoy, Douwe Kiela, Sarah Schwettmann,
  and Tristan Thrush.
\newblock Nearest neighbor normalization improves multimodal retrieval.
\newblock In \emph{Proceedings of the 2024 Conference on Empirical Methods in
  Natural Language Processing}, 2024.

\bibitem[Conti et~al.(2025)Conti, Faysse, Viaud, Bosselut, Hudelot, and
  Colombo]{insent2025}
Max Conti, Manuel Faysse, Gautier Viaud, Antoine Bosselut, Celine Hudelot, and
  Pierre Colombo.
\newblock Context is gold to find the gold passage: Evaluating and training
  contextual document embeddings.
\newblock In \emph{Proceedings of the 2025 Conference on Empirical Methods in
  Natural Language Processing}, 2025.

\bibitem[Dasigi et~al.(2021)Dasigi, Lo, Beltagy, Cohan, Smith, and
  Gardner]{qasper}
Pradeep Dasigi, Kyle Lo, Iz~Beltagy, Arman Cohan, Noah~A. Smith, and Matt
  Gardner.
\newblock A dataset of information-seeking questions and answers anchored in
  research papers.
\newblock In \emph{Proceedings of the 2021 Conference of the North American
  Chapter of the Association for Computational Linguistics: Human Language
  Technologies}, 2021.

\bibitem[Dhillon \& Modha(2001)Dhillon and Modha]{sphericalkmeans}
Inderjit~S. Dhillon and Dharmendra~S. Modha.
\newblock Concept decompositions for large sparse text data using clustering.
\newblock \emph{Machine Learning}, 42:\penalty0 143--175, 2001.

\bibitem[Esfandiarpoor et~al.(2026)Esfandiarpoor, Osmulski, Babakhin, de~Souza
  P.~Moreira, Holworthy, He, Ak, Cai, Chesler, Liu, and
  Oldridge]{esfandiarpoor2026semanticsimilarityperformancecosts}
Reza Esfandiarpoor, Radek Osmulski, Yauhen Babakhin, Gabriel de~Souza
  P.~Moreira, Oliver Holworthy, Jie He, Ronay Ak, Jiarui Cai, Ryan Chesler,
  Bo~Liu, and Even Oldridge.
\newblock Beyond semantic similarity: Performance and costs of agentic
  retrieval for complex tasks.
\newblock \emph{arXiv preprint arXiv:2610.05750}, 2026.

\bibitem[Fang et~al.(2026)Fang, Deng, Xu, Jiang, Tang, Xu, Deng, Yao, Wang,
  Qiao, Chen, and Zhang]{lightmem2026}
Jizhan Fang, Xinle Deng, Haoming Xu, Ziyan Jiang, Yuqi Tang, Ziwen Xu, Shumin
  Deng, Yunzhi Yao, Mengru Wang, Shuofei Qiao, Huajun Chen, and Ningyu Zhang.
\newblock {LightMem}: Lightweight and efficient memory-augmented generation.
\newblock In \emph{The Fourteenth International Conference on Learning
  Representations}, 2026.

\bibitem[Hara et~al.(2015)Hara, Suzuki, Shimbo, Kobayashi, Fukumizu, and
  Radovanovi{\'c}]{lc}
Kazuo Hara, Ikumi Suzuki, Masashi Shimbo, Kei Kobayashi, Kenji Fukumizu, and
  Milo{\v s} Radovanovi{\'c}.
\newblock Localized centering: Reducing hubness in large-sample data.
\newblock In \emph{Proceedings of the AAAI Conference on Artificial
  Intelligence}, 2015.

\bibitem[Jian \& Wang(2023)Jian and Wang]{invgc}
Xiangru Jian and Yimu Wang.
\newblock {InvGC}: Robust cross-modal retrieval by inverse graph convolution.
\newblock In \emph{Findings of the Association for Computational Linguistics:
  EMNLP 2023}, 2023.

\bibitem[Johnson et~al.(2021)Johnson, Douze, and J{\'e}gou]{faiss}
Jeff Johnson, Matthijs Douze, and Herv{\'e} J{\'e}gou.
\newblock Billion-scale similarity search with {GPU}s.
\newblock \emph{IEEE Transactions on Big Data}, 7\penalty0 (3):\penalty0
  535--547, 2021.

\bibitem[Kang et~al.(2025)Kang, Ji, Zhao, and Bai]{memoryos2025}
Jiazheng Kang, Mingming Ji, Zhe Zhao, and Ting Bai.
\newblock Memory {OS} of {AI} agent.
\newblock In \emph{Proceedings of the 2025 Conference on Empirical Methods in
  Natural Language Processing}, 2025.

\bibitem[Lee et~al.(2025)Lee, Maharana, Pujara, Ren, and Barbieri]{realtalk}
Dong-Ho Lee, Adyasha Maharana, Jay Pujara, Xiang Ren, and Francesco Barbieri.
\newblock Realtalk: A 21-day real-world dataset for long-term conversation.
\newblock \emph{arXiv preprint arXiv:2502.13270}, 2025.

\bibitem[Lee et~al.(2023)Lee, Dai, Duddu, Lei, Naim, Chang, and Zhao]{xtr}
Jinhyuk Lee, Zhuyun Dai, Sai Meher~Karthik Duddu, Tao Lei, Iftekhar Naim,
  Ming-Wei Chang, and Vincent Zhao.
\newblock Rethinking the role of token retrieval in multi-vector retrieval.
\newblock In \emph{Advances in Neural Information Processing Systems}, 2023.

\bibitem[Luo et~al.(2026)Luo, Wang, Chen, and Lei]{prq}
Yunxiao Luo, Siyuan Wang, Ben Chen, and Chenyi Lei.
\newblock Prq-kmeans: Projection residual quantization for semantic id
  tokenization.
\newblock \emph{arXiv preprint arXiv:2608.24207}, 2026.

\bibitem[Mac{\'e} et~al.(2025)Mac{\'e}, Loison, and Faysse]{vidorev2}
Quentin Mac{\'e}, Ant{\'o}nio Loison, and Manuel Faysse.
\newblock Vidore benchmark v2: Raising the bar for visual retrieval.
\newblock \emph{arXiv preprint arXiv:2505.17166}, 2025.

\bibitem[Maharana et~al.(2024)Maharana, Lee, Tulyakov, Bansal, Barbieri, and
  Fang]{locomo}
Adyasha Maharana, Dong-Ho Lee, Sergey Tulyakov, Mohit Bansal, Francesco
  Barbieri, and Yuwei Fang.
\newblock Evaluating very long-term conversational memory of {LLM} agents.
\newblock In \emph{Proceedings of the 62nd Annual Meeting of the Association
  for Computational Linguistics (Volume 1: Long Papers)}, 2024.

\bibitem[Malkov \& Yashunin(2020)Malkov and Yashunin]{hnsw}
Yury~A. Malkov and Dmitry~A. Yashunin.
\newblock Efficient and robust approximate nearest neighbor search using
  hierarchical navigable small world graphs.
\newblock \emph{IEEE Transactions on Pattern Analysis and Machine
  Intelligence}, 42\penalty0 (4):\penalty0 824--836, 2020.

\bibitem[M{\"o}ller et~al.(2020)M{\"o}ller, Reina, Jayakumar, and
  Pietsch]{covidqa}
Timo M{\"o}ller, Anthony Reina, Raghavan Jayakumar, and Malte Pietsch.
\newblock {COVID-QA}: A question answering dataset for {COVID}-19.
\newblock In \emph{Proceedings of the 1st Workshop on NLP for COVID-19 at ACL
  2020}, 2020.

\bibitem[Moreira et~al.(2024)Moreira, Osmulski, Xu, Ak, Schifferer, and
  Oldridge]{nvretriever}
Gabriel de Souza~P Moreira, Radek Osmulski, Mengyao Xu, Ronay Ak, Benedikt
  Schifferer, and Even Oldridge.
\newblock Nv-retriever: Improving text embedding models with effective
  hard-negative mining.
\newblock \emph{arXiv preprint arXiv:2407.15831}, 2024.

\bibitem[Morris \& Rush(2025)Morris and Rush]{cde}
John~X. Morris and Alexander~M. Rush.
\newblock Contextual document embeddings.
\newblock In \emph{The Thirteenth International Conference on Learning
  Representations}, 2025.

\bibitem[Nie et~al.(2026)Nie, Fu, Feng, and Shan]{evoembedding2026}
Chang Nie, Chaoyou Fu, Junlan Feng, and Caifeng Shan.
\newblock Evoembedding: Evolvable representations for long-context retrieval
  and agentic memory.
\newblock \emph{arXiv preprint arXiv:2606.21649}, 2026.

\bibitem[Nussbaum et~al.(2025)Nussbaum, Morris, Duderstadt, and
  Mulyar]{modernbert}
Zach Nussbaum, John~X. Morris, Brandon Duderstadt, and Andriy Mulyar.
\newblock {Nomic Embed}: Training a reproducible long context text embedder.
\newblock \emph{Transactions on Machine Learning Research}, 2025.

\bibitem[{OpenAI}(2024)]{gpt4omini}
{OpenAI}.
\newblock {GPT-4o mini}: Advancing cost-efficient intelligence.
\newblock
  \url{https://openai.com/index/gpt-4o-mini-advancing-cost-efficient-intelligence/},
  2024.

\bibitem[Pan et~al.(2025)Pan, Wang, Wu, Zhang, and Bu]{dbsn}
Zhengxin Pan, Haishuai Wang, Fangyu Wu, Peng Zhang, and Jiajun Bu.
\newblock Hubness reduction with dual bank sinkhorn normalization for
  cross-modal retrieval.
\newblock In \emph{Proceedings of the ACM International Conference on
  Multimedia}, 2025.

\bibitem[Robertson \& Zaragoza(2009)Robertson and Zaragoza]{bm25}
Stephen Robertson and Hugo Zaragoza.
\newblock The probabilistic relevance framework: {BM25} and beyond.
\newblock \emph{Foundations and Trends in Information Retrieval}, 3\penalty0
  (4):\penalty0 333--389, 2009.

\bibitem[Santhanam et~al.(2022{\natexlab{a}})Santhanam, Khattab, Potts, and
  Zaharia]{plaid}
Keshav Santhanam, Omar Khattab, Christopher Potts, and Matei Zaharia.
\newblock {PLAID}: An efficient engine for late interaction retrieval.
\newblock In \emph{Proceedings of the 31st ACM International Conference on
  Information \& Knowledge Management}, 2022{\natexlab{a}}.

\bibitem[Santhanam et~al.(2022{\natexlab{b}})Santhanam, Khattab, Saad-Falcon,
  Potts, and Zaharia]{colbertv2}
Keshav Santhanam, Omar Khattab, Jon Saad-Falcon, Christopher Potts, and Matei
  Zaharia.
\newblock {ColBERTv2}: Effective and efficient retrieval via lightweight late
  interaction.
\newblock In \emph{Proceedings of the 2022 Conference of the North American
  Chapter of the Association for Computational Linguistics: Human Language
  Technologies}, 2022{\natexlab{b}}.

\bibitem[Scheerer et~al.(2025)Scheerer, Zaharia, Potts, Alonso, and
  Khattab]{warp}
Jan~Luca Scheerer, Matei Zaharia, Christopher Potts, Gustavo Alonso, and Omar
  Khattab.
\newblock {WARP}: An efficient engine for multi-vector retrieval.
\newblock In \emph{Proceedings of the International ACM SIGIR Conference on
  Research and Development in Information Retrieval}, 2025.

\bibitem[Sun et~al.(2017)Sun, Li, Zhou, Tian, and Li]{lrs}
Shaoyan Sun, Ying Li, Wengang Zhou, Qi~Tian, and Houqiang Li.
\newblock Local residual similarity for image re-ranking.
\newblock \emph{Information Sciences}, 417:\penalty0 143--153, 2017.

\bibitem[Vera et~al.(2025)Vera, Dua, Zhang, Salz, Mullins, Panyam, Smoot, Naim,
  Zou, Chen, et~al.]{embeddinggemma}
Henrique~Schechter Vera, Sahil Dua, Biao Zhang, Daniel Salz, Ryan Mullins,
  Sindhu~Raghuram Panyam, Sara Smoot, Iftekhar Naim, Joe Zou, Feiyang Chen,
  et~al.
\newblock Embeddinggemma: Powerful and lightweight text representations.
\newblock \emph{arXiv preprint arXiv:2509.20354}, 2025.

\bibitem[Wu et~al.(2025)Wu, Wang, Yu, Zhang, Chang, and Yu]{longmem}
Di~Wu, Hongwei Wang, Wenhao Yu, Yuwei Zhang, Kai-Wei Chang, and Dong Yu.
\newblock {LongMemEval}: Benchmarking chat assistants on long-term interactive
  memory.
\newblock In \emph{The Thirteenth International Conference on Learning
  Representations}, 2025.

\bibitem[Xu et~al.(2025)Xu, Liang, Mei, Gao, Tan, and Zhang]{amem2025}
Wujiang Xu, Zujie Liang, Kai Mei, Hang Gao, Juntao Tan, and Yongfeng Zhang.
\newblock {A-Mem}: Agentic memory for {LLM} agents.
\newblock In \emph{Advances in Neural Information Processing Systems}, 2025.

\bibitem[Yang et~al.(2025)Yang, Li, Yang, Zhang, Hui, Zheng, Yu, Gao, Huang,
  Lv, et~al.]{qwen3}
An~Yang, Anfeng Li, Baosong Yang, Beichen Zhang, Binyuan Hui, Bo~Zheng, Bowen
  Yu, Chang Gao, Chengen Huang, Chenxu Lv, et~al.
\newblock Qwen3 technical report.
\newblock \emph{arXiv preprint arXiv:2505.09388}, 2025.

\bibitem[Yoon et~al.(2024)Yoon, Chen, Arik, and Pfister]{searchadaptor}
Jinsung Yoon, Yanfei Chen, Sercan Arik, and Tomas Pfister.
\newblock Search-adaptor: Embedding customization for information retrieval.
\newblock In \emph{Proceedings of the 62nd Annual Meeting of the Association
  for Computational Linguistics (Volume 1: Long Papers)}, 2024.

\bibitem[Zhang et~al.(2026)Zhang, Zhang, Ahmed, Jiang, Kumar, Sun, Lin, Sharma,
  Oraby, Colak, Aly, Kumar, Liu, and Dong]{assomem2026}
Kai Zhang, Xinyuan Zhang, Ejaz Ahmed, Hongda Jiang, Caleb Kumar, Kai Sun,
  Zhaojiang Lin, Sanat Sharma, Shereen Oraby, Aaron Colak, Ahmed Aly, Anuj
  Kumar, Xiaozhong Liu, and Xin Dong.
\newblock {AssoMem}: Scalable memory {QA} with multi-signal associative
  retrieval.
\newblock In \emph{The Fourteenth International Conference on Learning
  Representations}, 2026.

\bibitem[Zhang et~al.(2025)Zhang, Li, Long, Zhang, Lin, Yang, Xie, Yang, Liu,
  Lin, et~al.]{qwen}
Yanzhao Zhang, Mingxin Li, Dingkun Long, Xin Zhang, Huan Lin, Baosong Yang,
  Pengjun Xie, An~Yang, Dayiheng Liu, Junyang Lin, et~al.
\newblock Qwen3 embedding: Advancing text embedding and reranking through
  foundation models.
\newblock \emph{arXiv preprint arXiv:2506.05176}, 2025.

\bibitem[Zhao et~al.(2026{\natexlab{a}})Zhao, Hu, Shan, Huang, Zhou, Zhang,
  Sun, Liu, Li, Wei, Pan, Xiang, Zhang, Wang, Yu, Hu, and Zhang]{kalm}
Xinping Zhao, Xinshuo Hu, Zifei Shan, Shouzheng Huang, Yao Zhou, Xin Zhang,
  Zetian Sun, Zhenyu Liu, Dongfang Li, Xinyuan Wei, Youcheng Pan, Yang Xiang,
  Meishan Zhang, Haofen Wang, Jun Yu, Baotian Hu, and Min Zhang.
\newblock {KaLM-Embedding-V2}: Superior training techniques and data inspire a
  versatile embedding model.
\newblock In \emph{The Fourteenth International Conference on Learning
  Representations}, 2026{\natexlab{a}}.

\bibitem[Zhao et~al.(2026{\natexlab{b}})Zhao, Hu, Xu, Tang, Zhang, Zhou, Zhong,
  Zhou, Shan, Zhang, et~al.]{lmeb}
Xinping Zhao, Xinshuo Hu, Jiaxin Xu, Danyu Tang, Xin Zhang, Mengjia Zhou, Yan
  Zhong, Yao Zhou, Zifei Shan, Meishan Zhang, et~al.
\newblock Lmeb: Long-horizon memory embedding benchmark.
\newblock \emph{arXiv preprint arXiv:2603.12572}, 2026{\natexlab{b}}.

\end{thebibliography}
\bibliographystyle{near_references}
\endgroup
\clearpage
\appendix
\setlength{\intextsep}{6pt plus 1pt minus 1pt}
\captionsetup[table]{skip=5pt}
\setcounter{table}{0}
\renewcommand{\thetable}{A\arabic{table}}
\renewcommand{\theHtable}{appendix.\arabic{table}}
\setcounter{figure}{0}
\renewcommand{\thefigure}{A\arabic{figure}}
\renewcommand{\theHfigure}{appendix.\arabic{figure}}

\section{Evaluation details}
\label{app:population}
We evaluate 5,873 queries across eight benchmarks. Following LMEB, we report capped Recall@10 and NDCG@10 with linear relevance gains.
\begin{center}
\nearUnifiedTableStyle
\small
\setlength{\tabcolsep}{3pt}
\renewcommand{\arraystretch}{1.16}
\begin{tabular*}{\linewidth}{@{\hspace{\nearTableEdgePadding}\extracolsep{\fill}}lrrrr@{\hspace{\nearTableEdgePadding}}}
\toprule
\textbf{Dataset} & Queries & Documents & Banks & Subsets\\
\midrule
LoCoMo & 1,976 & 5,882 & 10 & 5\\
LongMemEval & 500 & 237,655 & 500 & 6\\
REALTALK & 679 & 8,944 & 10 & 3\\
QASPER & 1,335 & 20,508 & 413 & 1\\
ESG-Reports & 36 & 2,407 & 18 & 1\\
PeerQA & 136 & 18,593 & 70 & 1\\
CovidQA & 1,111 & 3,351 & 112 & 1\\
MLDR & 100 & 1,536 & 100 & 1\\
\bottomrule
\end{tabular*}
\end{center}

\noindent\textbf{Model configurations.}
Each encoder follows its released query/document format and pooling strategy. Same-encoder comparisons use identical inputs and cached embeddings.
The KaLM and ModernBERT citations follow their official model cards; the checkpoints used are \href{https://huggingface.co/tencent/KaLM-Embedding-Gemma3-12B-2511}{KaLM-Embedding-Gemma3-12B-2511} and \href{https://huggingface.co/nomic-ai/modernbert-embed-base}{ModernBERT-Embed-Base}.
The two smaller Llama models are \href{https://huggingface.co/nvidia/llama-nemotron-embed-1b-v2}{Llama-Nemotron-Embed-1B-v2} \citep{nvretriever} and \href{https://huggingface.co/nvidia/llama-nv-embed-reasoning-3b}{Llama-NV-Embed-Reasoning-3B} \citep{esfandiarpoor2026semanticsimilarityperformancecosts}.

\noindent\textbf{Baselines.}
LC, InvGC, and DBSN use the same frozen embeddings as NEAR, with configurations selected on aligned development splits. Search-Adaptor uses a shared residual MLP with hidden width 256, trained with Adam at learning rate $10^{-3}$ and batch size up to 128 for at most 2,000 steps \citep{searchadaptor}. ColBERTv2 and XTR retain their native encoders and input settings. We use PLAID \citep{plaid} for ColBERTv2 and WARP \citep{warp} for XTR. BM25 uses the default BM25Okapi parameters.

\noindent\textbf{Fixed settings.}
We fix the neighborhood width to 50, the reference-table size to 16, the number of routing entries to four, and the inverse temperature to 10. Depth, residual normalization, and total residual weight are selected on development data.

\noindent\textbf{Downstream question answering.}
\label{app:qa-protocol}
Retrieved memories are restored to chronological order before answer generation, which uses temperature zero and a maximum of 1,024 output tokens. Correctness is evaluated with fixed task-specific judge prompts. LoCoMo F1 follows its category-specific procedure; LongMemEval F1 measures normalized, stemmed word overlap on answerable questions. Base and NEAR are compared on the same questions.

\section{Development selection}
\label{app:selection}
For \ours{}, we search depth $L\in\{1,2,3\}$, raw or normalized residuals, and total residual weight in $\{0.25,0.5,1,2,4\}$. For seven benchmarks, we use grouped five-fold evaluation, assigning queries that share candidate banks or overlapping candidates to the same group. MLDR uses disjoint ConTEB queries for development. All methods use aligned splits and select their configurations by development NDCG@10. The selected configurations are reused for held-out Top-10 and Top-5 evaluation. Results for the seven grouped benchmarks are averaged over three repetitions.

\section{Implementation details}
\label{app:proof}
\label{sec:analysis}
For each matching representation $\ell=0,\ldots,L$, we compute a fixed score scale from the memory bank:
\begin{gather*}
\tau_\ell=\max\left\{10^{-6},
\sqrt{\frac{1}{N}\sum\nolimits_{i=1}^{N}
\operatorname{Var}_{j=1,\ldots,N}
\left[(\mathbf{z}_i^{(\ell)})^\top\mathbf{z}_j^{(\ell)}\right]}
\right\},
\end{gather*}
where $\operatorname{Var}$ is the population variance over all $N$ memories, including $j=i$.
The scales are computed offline without queries or relevance labels and shared by all queries in the same bank.
The implementation evaluates this expression using the covariance of the memory representations, without constructing all pairwise scores.

The query-reference table contains at most 16 entries and is constructed from the memory embeddings using spherical K-means. At retrieval time, the original and residual matching components are combined into a single inner-product representation, allowing \ours{} to use one ANN search over a fixed index.

For ANN retrieval, we use FAISS HNSW with inner-product search. We set $M=32$, \texttt{efConstruction} to 200, \texttt{efSearch} to 128, and the random seed to 42.

\section{Complete retrieval results}
\label{app:allencoders}
\begin{table}[H]
\centering
\caption{\textbf{Complete cross-model comparison (\%).} Eight-task mean capped Recall@10 and NDCG@10 for all same-backbone methods. Bold marks row bests, including ties.}
\label{tab:crossmodel-full}
\begingroup\nearUnifiedTableStyle\nearTableRules\fontsize{7.5}{7.6}\selectfont\setlength{\tabcolsep}{3pt}
\begin{tabular*}{\linewidth}{@{\hspace{\nearTableEdgePadding}\extracolsep{\fill}}lrrrrr>{\columncolor{nearTableTint}[\tabcolsep][\nearTableEdgePadding]}r@{\hspace{\nearTableEdgePadding}}}\toprule
\multicolumn{7}{c}{\textbf{(a) Capped Recall@10}}\\
\midrule
\textbf{Model} & \textbf{Base} & \textbf{LC} & \textbf{InvGC} & \textbf{DBSN} & \textbf{Search-Adaptor} & \textbf{NEAR}\\
\midrule
Qwen3-Embedding-0.6B & 65.34 & 65.74 & 65.67 & 65.42 & 66.42 & \textbf{67.17}\\
Qwen3-Embedding-4B & 62.67 & 62.91 & 68.44 & 69.17 & 67.97 & \textbf{69.76}\\
Qwen3-Embedding-8B & 68.65 & 69.32 & 69.35 & 70.08 & 69.73 & \textbf{71.76}\\
EmbeddingGemma-300M & 72.66 & 73.80 & 72.49 & 72.78 & 72.22 & \textbf{76.09}\\
KaLM-Embedding-Gemma3-12B & 72.67 & 73.44 & 72.29 & 73.33 & 73.51 & \textbf{75.19}\\
ModernBERT-Embed-Base & 69.76 & 70.00 & 69.26 & 69.42 & 69.53 & \textbf{70.88}\\
Llama-Nemotron-Embed-1B-v2 & 74.61 & 74.80 & 74.66 & 74.32 & 74.62 & \textbf{75.05}\\
Llama-NV-Embed-Reasoning-3B & 77.50 & 76.85 & 76.95 & 76.71 & 75.65 & \textbf{78.78}\\
Llama-Embed-Nemotron-8B & 76.51 & 76.67 & 76.19 & 76.24 & 76.67 & \textbf{77.88}\\
\midrule
\multicolumn{7}{c}{\textbf{(b) NDCG@10}}\\
\midrule
\textbf{Model} & \textbf{Base} & \textbf{LC} & \textbf{InvGC} & \textbf{DBSN} & \textbf{Search-Adaptor} & \textbf{NEAR}\\
\midrule
Qwen3-Embedding-0.6B & 50.93 & 51.69 & 51.67 & 51.15 & 52.23 & \textbf{52.99}\\
Qwen3-Embedding-4B & 47.88 & 48.50 & 54.36 & 54.31 & 54.40 & \textbf{55.33}\\
Qwen3-Embedding-8B & 53.95 & 54.23 & 54.64 & 55.37 & 55.50 & \textbf{57.00}\\
EmbeddingGemma-300M & 59.36 & 59.38 & 59.00 & 59.03 & 59.25 & \textbf{61.31}\\
KaLM-Embedding-Gemma3-12B & 57.11 & 57.52 & 57.75 & 58.21 & 58.80 & \textbf{59.21}\\
ModernBERT-Embed-Base & 55.72 & 55.78 & 55.71 & 55.41 & 55.45 & \textbf{57.08}\\
Llama-Nemotron-Embed-1B-v2 & 61.25 & 61.01 & 61.09 & 61.01 & 60.91 & \textbf{61.55}\\
Llama-NV-Embed-Reasoning-3B & 60.96 & 60.50 & 61.00 & 60.60 & 59.49 & \textbf{62.40}\\
Llama-Embed-Nemotron-8B & 62.73 & 62.47 & 62.48 & 62.47 & 63.03 & \textbf{63.71}\\
\bottomrule\end{tabular*}\endgroup

\end{table}
The tables report task-level Recall@10 and NDCG@10 for nine encoders and six same-encoder methods. The two smaller Llama-derived encoders are grouped for legibility.
\begin{table}[H]
\centering
\caption{\textbf{Additional Llama model results (\%).} Bold marks the best value within each model and column.}
\label{tab:llama-complete}
\begingroup\nearUnifiedTableStyle\neutralMainTableStyle\fontsize{7.2}{7.8}\selectfont\setlength{\tabcolsep}{2.05pt}
\begin{tabular*}{\linewidth}{@{\hspace{\nearTableEdgePadding}\extracolsep{\fill}}>{\columncolor{white}[\nearTableEdgePadding][\tabcolsep]}l*{8}{r}>{\columncolor{white}[\tabcolsep][\nearTableEdgePadding]}r@{\hspace{\nearTableEdgePadding}}}\toprule
\multicolumn{10}{c}{\textbf{(a) Capped Recall@10 $\uparrow$}}\\
\midrule
\textbf{Method} & \textbf{LoCoMo} & \textbf{LongMemEval} & \textbf{REALTALK} & \textbf{ESG-Reports} & \textbf{QASPER} & \textbf{PeerQA} & \textbf{CovidQA} & \textbf{MLDR} & \textbf{Overall}\\
\midrule
\rowcolor{nearGroupTint}\multicolumn{10}{c}{\textbf{Llama-Nemotron-Embed-1B-v2}}\\
Base & 68.33 & 93.60 & 56.37 & 60.60 & 77.48 & 45.84 & 95.68 & \textbf{99.00} & 74.61\\
LC & 68.40 & 93.69 & 56.71 & 60.83 & 77.46 & 46.69 & 95.65 & \textbf{99.00} & 74.80\,{\scriptsize$\pm$0.25}\\
InvGC & 68.86 & \textbf{93.91} & \textbf{57.07} & 59.10 & 77.47 & 45.76 & 96.07 & \textbf{99.00} & 74.66\,{\scriptsize$\pm$0.43}\\
DBSN & 68.04 & 93.60 & 56.61 & 59.06 & 77.48 & 45.13 & 95.68 & \textbf{99.00} & 74.32\,{\scriptsize$\pm$0.33}\\
Search-Adaptor & 69.00 & 93.75 & 56.89 & 55.77 & \textbf{78.81} & \textbf{47.64} & \textbf{96.10} & \textbf{99.00} & 74.62\,{\scriptsize$\pm$0.04}\\
\rowcolor{neutralTableGray}\textbf{NEAR} & \textbf{69.53} & 93.45 & 56.13 & \textbf{61.88} & 77.74 & 46.97 & 95.71 & \textbf{99.00} & \textbf{75.05}\,{\scriptsize$\pm$0.36}\\
\midrule
\rowcolor{nearGroupTint}\multicolumn{10}{c}{\textbf{Llama-NV-Embed-Reasoning-3B}}\\
Base & 72.09 & 89.39 & 61.20 & 78.98 & 80.78 & 44.97 & 93.61 & \textbf{99.00} & 77.50\\
LC & 71.92 & 89.23 & 61.20 & 73.77 & 80.70 & 45.37 & 93.58 & \textbf{99.00} & 76.85\,{\scriptsize$\pm$0.26}\\
InvGC & 72.02 & 88.83 & 61.61 & 74.24 & 80.71 & 45.26 & \textbf{93.91} & \textbf{99.00} & 76.95\,{\scriptsize$\pm$0.45}\\
DBSN & 72.09 & 89.56 & 60.84 & 73.52 & 80.78 & 44.39 & 93.55 & \textbf{99.00} & 76.71\,{\scriptsize$\pm$0.11}\\
Search-Adaptor & 71.87 & 90.08 & 61.22 & 64.65 & \textbf{81.57} & 43.66 & 93.19 & \textbf{99.00} & 75.65\,{\scriptsize$\pm$0.81}\\
\rowcolor{neutralTableGray}\textbf{NEAR} & \textbf{74.42} & \textbf{90.52} & \textbf{61.73} & \textbf{83.66} & 80.23 & \textbf{47.02} & 93.64 & \textbf{99.00} & \textbf{78.78}\,{\scriptsize$\pm$0.10}\\
\bottomrule\end{tabular*}\endgroup
\par\vspace{3pt}
\begingroup\nearUnifiedTableStyle\neutralMainTableStyle\fontsize{7.2}{7.8}\selectfont\setlength{\tabcolsep}{2.05pt}
\begin{tabular*}{\linewidth}{@{\hspace{\nearTableEdgePadding}\extracolsep{\fill}}>{\columncolor{white}[\nearTableEdgePadding][\tabcolsep]}l*{8}{r}>{\columncolor{white}[\tabcolsep][\nearTableEdgePadding]}r@{\hspace{\nearTableEdgePadding}}}\toprule
\multicolumn{10}{c}{\textbf{(b) NDCG@10 $\uparrow$}}\\
\midrule
\textbf{Method} & \textbf{LoCoMo} & \textbf{LongMemEval} & \textbf{REALTALK} & \textbf{ESG-Reports} & \textbf{QASPER} & \textbf{PeerQA} & \textbf{CovidQA} & \textbf{MLDR} & \textbf{Overall}\\
\midrule
\rowcolor{nearGroupTint}\multicolumn{10}{c}{\textbf{Llama-Nemotron-Embed-1B-v2}}\\
Base & 53.88 & 83.04 & 47.42 & 48.21 & 56.32 & \textbf{36.31} & 84.56 & \textbf{80.26} & 61.25\\
LC & 54.31 & 83.19 & 47.28 & 47.05 & 56.06 & 35.58 & 84.36 & \textbf{80.26} & 61.01\,{\scriptsize$\pm$0.03}\\
InvGC & 54.21 & 84.10 & \textbf{48.27} & 46.01 & 56.22 & 35.14 & 84.51 & 80.24 & 61.09\,{\scriptsize$\pm$0.22}\\
DBSN & 53.58 & 83.97 & 47.38 & 47.13 & 56.32 & 34.91 & 84.56 & \textbf{80.26} & 61.01\,{\scriptsize$\pm$0.16}\\
Search-Adaptor & 54.43 & 83.14 & 47.88 & 42.27 & \textbf{58.73} & 35.83 & \textbf{84.76} & \textbf{80.26} & 60.91\,{\scriptsize$\pm$0.20}\\
\rowcolor{neutralTableGray}\textbf{NEAR} & \textbf{55.22} & \textbf{84.45} & 47.06 & \textbf{48.52} & 56.32 & 36.00 & 84.56 & \textbf{80.26} & \textbf{61.55}\,{\scriptsize$\pm$0.27}\\
\midrule
\rowcolor{nearGroupTint}\multicolumn{10}{c}{\textbf{Llama-NV-Embed-Reasoning-3B}}\\
Base & 56.13 & 75.51 & 49.19 & 56.74 & 58.12 & 34.23 & 76.69 & 81.10 & 60.96\\
LC & 55.98 & 75.34 & 49.02 & 53.63 & 58.01 & 34.29 & 76.66 & 81.10 & 60.50\,{\scriptsize$\pm$0.23}\\
InvGC & 56.24 & 75.73 & \textbf{49.63} & 55.30 & 58.18 & 33.68 & 77.20 & \textbf{82.07} & 61.00\,{\scriptsize$\pm$0.18}\\
DBSN & 56.13 & 76.43 & 48.70 & 54.50 & 58.12 & 33.28 & 76.52 & 81.10 & 60.60\,{\scriptsize$\pm$0.46}\\
Search-Adaptor & 56.14 & 75.82 & 48.96 & 42.87 & \textbf{60.13} & 32.80 & \textbf{77.38} & 81.84 & 59.49\,{\scriptsize$\pm$1.01}\\
\rowcolor{neutralTableGray}\textbf{NEAR} & \textbf{58.20} & \textbf{76.83} & 49.60 & \textbf{61.37} & 58.33 & \textbf{37.23} & 76.54 & 81.11 & \textbf{62.40}\,{\scriptsize$\pm$0.13}\\
\bottomrule\end{tabular*}\endgroup

\end{table}
\begin{table}[H]
\centering
\caption{\textbf{Complete capped Recall@10 results (\%).} Table~\ref{tab:llama-complete} adds the 1B and 3B Llama models. Overall reports the mean and sample SD across the three splits.}
\begingroup\nearUnifiedTableStyle\neutralMainTableStyle\fontsize{7.2}{8}\selectfont\setlength{\tabcolsep}{2.05pt}\renewcommand{\arraystretch}{1.23}
\begin{tabular*}{\linewidth}{@{\hspace{\nearTableEdgePadding}\extracolsep{\fill}}>{\columncolor{white}[\nearTableEdgePadding][\tabcolsep]}l*{8}{r}>{\columncolor{white}[\tabcolsep][\nearTableEdgePadding]}r@{\hspace{\nearTableEdgePadding}}}\toprule
\multicolumn{10}{c}{\textbf{(a) Capped Recall@10 $\uparrow$}}\\
\midrule
\textbf{Method} & \textbf{LoCoMo} & \textbf{LongMemEval} & \textbf{REALTALK} & \textbf{ESG-Reports} & \textbf{QASPER} & \textbf{PeerQA} & \textbf{CovidQA} & \textbf{MLDR} & \textbf{Overall}\\
\midrule
\rowcolor{nearGroupTint}\multicolumn{10}{c}{\textbf{Standalone retrieval systems}}\\
ColBERTv2 & 62.95 & 87.53 & 54.34 & 56.94 & 75.46 & 45.25 & 95.23 & 96.00 & 71.71\\
XTR & 43.55 & 89.42 & 42.65 & 62.87 & 71.19 & 38.82 & 93.88 & 98.00 & 67.55\\
\midrule
\rowcolor{nearGroupTint}\multicolumn{10}{c}{\textbf{Qwen3-Embedding-0.6B}}\\
Base & 46.44 & 84.84 & 46.39 & \textbf{52.82} & 69.71 & 36.17 & 91.36 & \textbf{95.00} & 65.34\\
LC & 50.38 & 85.87 & 51.06 & 46.13 & 69.55 & 37.11 & 90.85 & \textbf{95.00} & 65.74\,{\scriptsize$\pm$0.39}\\
InvGC & 48.10 & \textbf{88.70} & 49.27 & 45.62 & 69.79 & 36.69 & 92.23 & \textbf{95.00} & 65.67\,{\scriptsize$\pm$0.27}\\
DBSN & 46.55 & 86.72 & 46.71 & 48.07 & 69.71 & 38.08 & 92.53 & \textbf{95.00} & 65.42\,{\scriptsize$\pm$0.84}\\
Search-Adaptor & 51.88 & 88.45 & 49.31 & 41.40 & \textbf{73.69} & 38.57 & \textbf{93.07} & \textbf{95.00} & 66.42\,{\scriptsize$\pm$0.27}\\
\rowcolor{neutralTableGray}\textbf{NEAR} & \textbf{55.75} & 86.24 & \textbf{51.24} & 49.41 & 69.43 & \textbf{39.75} & 90.55 & \textbf{95.00} & \textbf{67.17}\,{\scriptsize$\pm$0.21}\\
\midrule
\rowcolor{nearGroupTint}\multicolumn{10}{c}{\textbf{Qwen3-Embedding-4B}}\\
Base & 41.39 & 58.96 & 42.47 & 60.69 & 73.11 & 35.19 & 93.52 & 96.00 & 62.67\\
LC & 44.59 & 58.96 & 44.12 & 56.19 & 72.97 & 37.36 & 93.07 & 96.00 & 62.91\,{\scriptsize$\pm$0.35}\\
InvGC & 49.31 & 91.33 & 46.77 & 59.06 & 73.25 & 38.34 & 93.49 & 96.00 & 68.44\,{\scriptsize$\pm$0.77}\\
DBSN & 51.71 & 91.69 & \textbf{49.51} & 58.41 & 73.11 & \textbf{39.58} & 93.37 & 96.00 & 69.17\,{\scriptsize$\pm$0.29}\\
Search-Adaptor & 50.07 & \textbf{93.57} & 43.71 & 52.90 & \textbf{75.65} & 37.88 & \textbf{94.00} & 96.00 & 67.97\,{\scriptsize$\pm$1.17}\\
\rowcolor{neutralTableGray}\textbf{NEAR} & \textbf{56.15} & 89.65 & 47.17 & \textbf{62.73} & 72.76 & 38.87 & 93.79 & \textbf{97.00} & \textbf{69.76}\,{\scriptsize$\pm$0.28}\\
\midrule
\rowcolor{nearGroupTint}\multicolumn{10}{c}{\textbf{Qwen3-Embedding-8B}}\\
Base & 49.15 & 87.93 & 49.78 & 62.22 & 70.38 & 40.76 & 93.97 & 95.00 & 68.65\\
LC & 50.66 & 87.89 & 50.50 & 63.12 & 70.27 & 42.12 & 93.97 & \textbf{96.00} & 69.32\,{\scriptsize$\pm$0.35}\\
InvGC & 51.97 & \textbf{94.74} & 51.93 & 56.20 & 71.62 & 38.18 & 94.15 & \textbf{96.00} & 69.35\,{\scriptsize$\pm$0.23}\\
DBSN & 51.75 & 94.69 & 51.34 & 62.33 & 70.38 & 41.43 & 93.70 & 95.00 & 70.08\,{\scriptsize$\pm$0.23}\\
Search-Adaptor & 54.65 & 93.07 & 50.57 & 52.85 & \textbf{74.92} & 41.18 & \textbf{94.57} & \textbf{96.00} & 69.73\,{\scriptsize$\pm$0.50}\\
\rowcolor{neutralTableGray}\textbf{NEAR} & \textbf{59.88} & 90.70 & \textbf{52.56} & \textbf{68.01} & 69.85 & \textbf{43.15} & 93.91 & \textbf{96.00} & \textbf{71.76}\,{\scriptsize$\pm$0.16}\\
\midrule
\rowcolor{nearGroupTint}\multicolumn{10}{c}{\textbf{EmbeddingGemma-300M}}\\
Base & 64.66 & \textbf{91.88} & 58.71 & 50.46 & 78.64 & 46.41 & 93.52 & 97.00 & 72.66\\
LC & 65.72 & 91.73 & 59.91 & 56.57 & 78.39 & 45.32 & 93.73 & 99.00 & 73.80\,{\scriptsize$\pm$0.35}\\
InvGC & 63.76 & 91.73 & 60.61 & 47.69 & 78.53 & 46.70 & 93.88 & 97.00 & 72.49\,{\scriptsize$\pm$0.39}\\
DBSN & 64.66 & 91.23 & 58.84 & 51.88 & 78.64 & 46.63 & 93.40 & 97.00 & 72.78\,{\scriptsize$\pm$0.38}\\
Search-Adaptor & 65.03 & 91.36 & 59.24 & 44.34 & \textbf{79.73} & \textbf{47.06} & 94.03 & 97.00 & 72.22\,{\scriptsize$\pm$0.86}\\
\rowcolor{neutralTableGray}\textbf{NEAR} & \textbf{70.38} & 91.76 & \textbf{60.85} & \textbf{68.09} & 78.43 & 45.09 & \textbf{94.12} & \textbf{100.00} & \textbf{76.09}\,{\scriptsize$\pm$0.41}\\
\midrule
\rowcolor{nearGroupTint}\multicolumn{10}{c}{\textbf{KaLM-Embedding-Gemma3-12B}}\\
Base & 58.50 & 93.07 & 54.99 & 63.61 & 73.46 & 44.78 & 92.98 & \textbf{100.00} & 72.67\\
LC & 61.60 & 92.89 & \textbf{57.47} & 64.29 & 73.46 & 45.07 & 92.74 & \textbf{100.00} & 73.44\,{\scriptsize$\pm$0.18}\\
InvGC & 60.34 & 92.85 & 55.68 & 58.73 & 73.84 & 44.24 & 93.67 & 99.00 & 72.29\,{\scriptsize$\pm$0.21}\\
DBSN & 61.09 & 93.56 & 55.04 & 64.12 & 73.46 & \textbf{47.22} & 93.16 & 99.00 & 73.33\,{\scriptsize$\pm$0.16}\\
Search-Adaptor & 64.76 & \textbf{94.85} & 56.76 & 55.12 & \textbf{76.88} & 45.36 & \textbf{94.39} & \textbf{100.00} & 73.51\,{\scriptsize$\pm$0.94}\\
\rowcolor{neutralTableGray}\textbf{NEAR} & \textbf{66.59} & 93.29 & 56.97 & \textbf{72.65} & 73.86 & 45.08 & 93.10 & \textbf{100.00} & \textbf{75.19}\,{\scriptsize$\pm$0.17}\\
\midrule
\rowcolor{nearGroupTint}\multicolumn{10}{c}{\textbf{ModernBERT-Embed-Base}}\\
Base & 58.93 & 87.40 & 51.44 & \textbf{51.76} & 76.57 & 42.40 & 92.62 & \textbf{97.00} & 69.76\\
LC & 61.24 & 87.23 & 53.30 & 48.80 & 76.40 & 43.54 & 92.50 & \textbf{97.00} & 70.00\,{\scriptsize$\pm$0.24}\\
InvGC & 58.30 & 88.47 & 52.43 & 46.90 & 76.35 & 42.01 & 92.59 & \textbf{97.00} & 69.26\,{\scriptsize$\pm$0.41}\\
DBSN & 58.93 & 86.61 & 51.14 & 50.56 & 76.57 & 41.97 & 92.62 & \textbf{97.00} & 69.42\,{\scriptsize$\pm$0.25}\\
Search-Adaptor & 60.36 & \textbf{88.54} & 52.14 & 46.16 & \textbf{77.21} & 42.15 & \textbf{92.68} & \textbf{97.00} & 69.53\,{\scriptsize$\pm$0.87}\\
\rowcolor{neutralTableGray}\textbf{NEAR} & \textbf{66.06} & 87.21 & \textbf{53.82} & 50.93 & 76.10 & \textbf{44.38} & 92.53 & 96.00 & \textbf{70.88}\,{\scriptsize$\pm$0.40}\\
\midrule
\rowcolor{nearGroupTint}\multicolumn{10}{c}{\textbf{Llama-Embed-Nemotron-8B}}\\
Base & 71.90 & 95.11 & 60.73 & 67.08 & 77.74 & 46.45 & 95.05 & 98.00 & 76.51\\
LC & 71.63 & 94.62 & 61.46 & 68.35 & 77.74 & 47.11 & 94.45 & 98.00 & 76.67\,{\scriptsize$\pm$0.21}\\
InvGC & 71.71 & \textbf{95.62} & 61.10 & 64.78 & 77.70 & 45.36 & 95.26 & 98.00 & 76.19\,{\scriptsize$\pm$0.28}\\
DBSN & 71.90 & 94.80 & 60.51 & 65.20 & 77.74 & 45.04 & \textbf{95.71} & \textbf{99.00} & 76.24\,{\scriptsize$\pm$0.58}\\
Search-Adaptor & 73.10 & 95.13 & 62.10 & 63.38 & \textbf{78.89} & 47.18 & 95.56 & 98.00 & 76.67\,{\scriptsize$\pm$1.10}\\
\rowcolor{neutralTableGray}\textbf{NEAR} & \textbf{73.60} & 94.82 & \textbf{62.37} & \textbf{72.62} & 77.91 & \textbf{47.97} & 94.78 & \textbf{99.00} & \textbf{77.88}\,{\scriptsize$\pm$0.17}\\
\bottomrule\end{tabular*}\endgroup

\end{table}
\begin{table}[H]
\centering
\caption{\textbf{Complete NDCG@10 results (\%).} The evaluation settings match Table~\ref{tab:main}.}
\begingroup\nearUnifiedTableStyle\neutralMainTableStyle\fontsize{7.2}{8}\selectfont\setlength{\tabcolsep}{2.05pt}\renewcommand{\arraystretch}{1.23}
\begin{tabular*}{\linewidth}{@{\hspace{\nearTableEdgePadding}\extracolsep{\fill}}>{\columncolor{white}[\nearTableEdgePadding][\tabcolsep]}l*{8}{r}>{\columncolor{white}[\tabcolsep][\nearTableEdgePadding]}r@{\hspace{\nearTableEdgePadding}}}\toprule
\multicolumn{10}{c}{\textbf{(b) NDCG@10 $\uparrow$}}\\
\midrule
\textbf{Method} & \textbf{LoCoMo} & \textbf{LongMemEval} & \textbf{REALTALK} & \textbf{ESG-Reports} & \textbf{QASPER} & \textbf{PeerQA} & \textbf{CovidQA} & \textbf{MLDR} & \textbf{Overall}\\
\midrule
\rowcolor{nearGroupTint}\multicolumn{10}{c}{\textbf{Standalone retrieval systems}}\\
ColBERTv2 & 49.97 & 79.98 & 45.81 & 45.05 & 51.97 & 36.29 & 84.44 & 85.35 & 59.86\\
XTR & 31.59 & 79.12 & 33.11 & 51.91 & 48.76 & 29.80 & 83.25 & 88.18 & 55.71\\
\midrule
\rowcolor{nearGroupTint}\multicolumn{10}{c}{\textbf{Qwen3-Embedding-0.6B}}\\
Base & 33.92 & 72.82 & 37.45 & 40.08 & 47.02 & 26.69 & 73.11 & 76.40 & 50.93\\
LC & 37.35 & 73.85 & \textbf{41.05} & 36.39 & 47.21 & 27.33 & 73.49 & \textbf{76.87} & 51.69\,{\scriptsize$\pm$0.22}\\
InvGC & 35.05 & 77.03 & 39.74 & 36.24 & 47.07 & 27.39 & 74.51 & 76.35 & 51.67\,{\scriptsize$\pm$0.22}\\
DBSN & 34.09 & 75.89 & 37.66 & 35.33 & 47.02 & 28.43 & 74.41 & 76.40 & 51.15\,{\scriptsize$\pm$0.38}\\
Search-Adaptor & 38.01 & \textbf{77.29} & 39.28 & 30.76 & \textbf{51.55} & 28.24 & \textbf{76.05} & 76.68 & 52.23\,{\scriptsize$\pm$0.13}\\
\rowcolor{neutralTableGray}\textbf{NEAR} & \textbf{41.46} & 75.14 & 40.82 & \textbf{40.48} & 47.37 & \textbf{29.03} & 73.59 & 76.07 & \textbf{52.99}\,{\scriptsize$\pm$0.27}\\
\midrule
\rowcolor{nearGroupTint}\multicolumn{10}{c}{\textbf{Qwen3-Embedding-4B}}\\
Base & 29.69 & 47.11 & 34.46 & 41.21 & 50.23 & 26.74 & 76.56 & 77.02 & 47.88\\
LC & 32.21 & 47.11 & 34.75 & 41.47 & 50.08 & 28.03 & 76.68 & 77.70 & 48.50\,{\scriptsize$\pm$0.34}\\
InvGC & 35.65 & 81.22 & 37.60 & 44.69 & 51.22 & 27.92 & 78.09 & \textbf{78.49} & 54.36\,{\scriptsize$\pm$0.67}\\
DBSN & 38.14 & 80.46 & \textbf{40.20} & 42.64 & 50.23 & 29.10 & 76.68 & 77.02 & 54.31\,{\scriptsize$\pm$0.31}\\
Search-Adaptor & 37.38 & \textbf{83.06} & 35.51 & 38.34 & \textbf{54.76} & 28.84 & \textbf{79.06} & 78.26 & 54.40\,{\scriptsize$\pm$0.62}\\
\rowcolor{neutralTableGray}\textbf{NEAR} & \textbf{42.04} & 78.50 & 37.70 & \textbf{47.76} & 50.90 & \textbf{30.30} & 77.63 & 77.83 & \textbf{55.33}\,{\scriptsize$\pm$0.25}\\
\midrule
\rowcolor{nearGroupTint}\multicolumn{10}{c}{\textbf{Qwen3-Embedding-8B}}\\
Base & 36.17 & 77.40 & 40.66 & 43.00 & 47.35 & 30.11 & 79.46 & 77.40 & 53.95\\
LC & 37.13 & 77.28 & 40.22 & 44.22 & 47.26 & 30.22 & 79.32 & 78.15 & 54.23\,{\scriptsize$\pm$0.17}\\
InvGC & 38.75 & 84.24 & \textbf{42.39} & 37.93 & 48.28 & 28.56 & \textbf{79.95} & 77.03 & 54.64\,{\scriptsize$\pm$0.05}\\
DBSN & 39.21 & \textbf{84.71} & 42.30 & 42.38 & 47.35 & 30.66 & 78.94 & 77.40 & 55.37\,{\scriptsize$\pm$0.06}\\
Search-Adaptor & 41.29 & 83.00 & 41.04 & 37.19 & \textbf{52.06} & 30.83 & 79.94 & \textbf{78.69} & 55.50\,{\scriptsize$\pm$0.38}\\
\rowcolor{neutralTableGray}\textbf{NEAR} & \textbf{45.34} & 81.64 & 42.33 & \textbf{49.86} & 47.13 & \textbf{32.09} & 79.15 & 78.48 & \textbf{57.00}\,{\scriptsize$\pm$0.17}\\
\midrule
\rowcolor{nearGroupTint}\multicolumn{10}{c}{\textbf{EmbeddingGemma-300M}}\\
Base & 50.29 & 79.81 & 48.47 & 42.33 & 56.85 & 37.49 & 78.79 & 80.88 & 59.36\\
LC & 51.12 & 79.70 & 48.99 & 43.08 & 56.84 & 36.37 & 78.55 & 80.37 & 59.38\,{\scriptsize$\pm$0.12}\\
InvGC & 49.94 & 80.14 & \textbf{49.66} & 38.49 & 56.70 & 37.34 & 78.84 & 80.88 & 59.00\,{\scriptsize$\pm$0.41}\\
DBSN & 50.29 & 79.76 & 48.32 & 40.03 & 56.85 & 37.47 & 78.63 & 80.88 & 59.03\,{\scriptsize$\pm$0.11}\\
Search-Adaptor & 51.05 & 79.87 & 48.85 & 35.27 & \textbf{59.85} & \textbf{38.21} & \textbf{78.86} & \textbf{82.05} & 59.25\,{\scriptsize$\pm$0.62}\\
\rowcolor{neutralTableGray}\textbf{NEAR} & \textbf{55.01} & \textbf{80.45} & 49.46 & \textbf{52.18} & 56.91 & 37.23 & 78.70 & 80.56 & \textbf{61.31}\,{\scriptsize$\pm$0.26}\\
\midrule
\rowcolor{nearGroupTint}\multicolumn{10}{c}{\textbf{KaLM-Embedding-Gemma3-12B}}\\
Base & 42.58 & 81.64 & 44.66 & 49.51 & 51.56 & 32.56 & 77.98 & 76.37 & 57.11\\
LC & 45.00 & 81.53 & 45.55 & 50.22 & 51.34 & 32.47 & 77.68 & 76.37 & 57.52\,{\scriptsize$\pm$0.04}\\
InvGC & 44.18 & 81.92 & 44.89 & 46.50 & 52.03 & 34.03 & 78.31 & 80.12 & 57.75\,{\scriptsize$\pm$0.22}\\
DBSN & 44.93 & 82.78 & 44.34 & 48.05 & 51.56 & \textbf{35.82} & 77.59 & \textbf{80.65} & 58.21\,{\scriptsize$\pm$0.28}\\
Search-Adaptor & \textbf{49.60} & \textbf{83.37} & \textbf{46.03} & 41.55 & \textbf{55.69} & 34.44 & \textbf{80.02} & 79.65 & 58.80\,{\scriptsize$\pm$0.36}\\
\rowcolor{neutralTableGray}\textbf{NEAR} & 49.43 & 81.74 & 45.53 & \textbf{54.78} & 51.69 & 34.67 & 77.87 & 78.00 & \textbf{59.21}\,{\scriptsize$\pm$0.31}\\
\midrule
\rowcolor{nearGroupTint}\multicolumn{10}{c}{\textbf{ModernBERT-Embed-Base}}\\
Base & 45.80 & 74.36 & 42.47 & 35.98 & 54.06 & \textbf{34.61} & \textbf{76.93} & 81.53 & 55.72\\
LC & 47.15 & 74.45 & 43.41 & 35.45 & 53.75 & 34.24 & 76.58 & 81.19 & 55.78\,{\scriptsize$\pm$0.14}\\
InvGC & 45.44 & \textbf{76.59} & 43.77 & 33.86 & 53.98 & 33.61 & 76.92 & 81.53 & 55.71\,{\scriptsize$\pm$0.39}\\
DBSN & 45.80 & 74.60 & 42.28 & 34.44 & 54.06 & 33.65 & \textbf{76.93} & 81.53 & 55.41\,{\scriptsize$\pm$0.22}\\
Search-Adaptor & 46.89 & 76.30 & 43.03 & 29.76 & \textbf{55.34} & 33.39 & 76.76 & \textbf{82.13} & 55.45\,{\scriptsize$\pm$0.76}\\
\rowcolor{neutralTableGray}\textbf{NEAR} & \textbf{51.46} & 74.92 & \textbf{44.08} & \textbf{39.68} & 53.87 & 34.46 & 76.91 & 81.28 & \textbf{57.08}\,{\scriptsize$\pm$0.12}\\
\midrule
\rowcolor{nearGroupTint}\multicolumn{10}{c}{\textbf{Llama-Embed-Nemotron-8B}}\\
Base & 57.69 & 85.84 & 51.13 & 50.94 & 56.55 & 35.55 & 80.98 & 83.16 & 62.73\\
LC & 57.19 & 85.51 & 51.44 & 50.46 & 56.55 & 35.04 & 80.65 & 82.94 & 62.47\,{\scriptsize$\pm$0.09}\\
InvGC & 57.52 & \textbf{86.17} & 51.13 & 48.86 & 56.41 & 34.67 & 81.56 & \textbf{83.53} & 62.48\,{\scriptsize$\pm$0.02}\\
DBSN & 57.69 & 85.61 & 50.98 & 49.85 & 56.55 & 34.42 & 81.64 & 83.03 & 62.47\,{\scriptsize$\pm$0.25}\\
Search-Adaptor & 59.13 & 85.71 & 51.93 & 46.88 & \textbf{58.71} & 36.25 & \textbf{82.48} & 83.16 & 63.03\,{\scriptsize$\pm$0.72}\\
\rowcolor{neutralTableGray}\textbf{NEAR} & \textbf{60.05} & 85.80 & \textbf{52.03} & \textbf{55.11} & 56.61 & \textbf{36.46} & 81.31 & 82.35 & \textbf{63.71}\,{\scriptsize$\pm$0.14}\\
\bottomrule\end{tabular*}\endgroup

\end{table}
\FloatBarrier

\clearpage
\section{Top-5 results}
\label{app:aggregate}
We report Top-5 results using the configurations selected for Top-10 evaluation.
\begin{table}[H]
\centering
\caption{\textbf{Top-5 retrieval results (\%).} Each cell reports capped Recall@5 above NDCG@5, averaged across eight benchmarks. Values are mean $\pm$ SD over three splits. Bold marks the best mean, including ties.}
\begingroup\nearUnifiedTableStyle\fontsize{7.8}{9.5}\selectfont\setlength{\tabcolsep}{1.7pt}
\begin{tabular*}{\linewidth}{@{\hspace{\nearTableEdgePadding}\extracolsep{\fill}}lrrrrr>{\columncolor{white}[\tabcolsep][\nearTableEdgePadding]}r@{\hspace{\nearTableEdgePadding}}}\toprule
\textbf{Model} & Base & LC & InvGC & DBSN & Search-Adaptor & \textbf{NEAR}\\\midrule
Qwen3-Embedding-0.6B & \shortstack{55.76\\47.40} & \shortstack{57.19\sdsmall{0.30}\\48.51\sdsmall{0.17}} & \shortstack{56.89\sdsmall{0.28}\\48.40\sdsmall{0.31}} & \shortstack{56.29\sdsmall{0.68}\\47.78\sdsmall{0.27}} & \shortstack{57.68\sdsmall{0.52}\\48.95\sdsmall{0.30}} & \cellcolor{oursblue}\shortstack{\textbf{57.80}\sdsmall{0.21}\\\textbf{49.49}\sdsmall{0.37}}\\[3pt]
Qwen3-Embedding-4B & \shortstack{52.18\\44.03} & \shortstack{53.00\sdsmall{0.40}\\44.87\sdsmall{0.39}} & \shortstack{59.34\sdsmall{0.35}\\50.94\sdsmall{0.65}} & \shortstack{59.66\sdsmall{0.23}\\50.74\sdsmall{0.27}} & \shortstack{58.35\sdsmall{0.42}\\50.82\sdsmall{0.37}} & \cellcolor{oursblue}\shortstack{\textbf{60.01}\sdsmall{0.33}\\\textbf{51.73}\sdsmall{0.31}}\\[3pt]
Qwen3-Embedding-8B & \shortstack{59.73\\50.57} & \shortstack{60.05\sdsmall{0.19}\\50.76\sdsmall{0.12}} & \shortstack{59.76\sdsmall{0.44}\\51.09\sdsmall{0.03}} & \shortstack{59.93\sdsmall{0.24}\\51.56\sdsmall{0.05}} & \shortstack{60.70\sdsmall{0.40}\\52.09\sdsmall{0.34}} & \cellcolor{oursblue}\shortstack{\textbf{62.84}\sdsmall{0.30}\\\textbf{53.60}\sdsmall{0.20}}\\[3pt]
EmbeddingGemma-300M & \shortstack{63.74\\56.02} & \shortstack{63.98\sdsmall{0.32}\\55.70\sdsmall{0.06}} & \shortstack{63.78\sdsmall{0.23}\\55.74\sdsmall{0.31}} & \shortstack{64.04\sdsmall{0.44}\\55.75\sdsmall{0.19}} & \shortstack{63.58\sdsmall{0.72}\\55.96\sdsmall{0.54}} & \cellcolor{oursblue}\shortstack{\textbf{66.72}\sdsmall{0.17}\\\textbf{57.83}\sdsmall{0.11}}\\[3pt]
KaLM-Embedding-Gemma3-12B & \shortstack{63.34\\53.56} & \shortstack{64.20\sdsmall{0.43}\\53.98\sdsmall{0.18}} & \shortstack{63.48\sdsmall{0.51}\\54.40\sdsmall{0.38}} & \shortstack{64.03\sdsmall{0.27}\\54.69\sdsmall{0.37}} & \shortstack{64.86\sdsmall{0.73}\\\textbf{55.49}\sdsmall{0.31}} & \cellcolor{oursblue}\shortstack{\textbf{65.34}\sdsmall{0.53}\\55.46\sdsmall{0.41}}\\[3pt]
ModernBERT-Embed-Base & \shortstack{59.32\\51.88} & \shortstack{60.48\sdsmall{0.91}\\52.27\sdsmall{0.36}} & \shortstack{60.10\sdsmall{0.43}\\52.36\sdsmall{0.27}} & \shortstack{59.41\sdsmall{0.14}\\51.73\sdsmall{0.19}} & \shortstack{60.30\sdsmall{0.89}\\52.05\sdsmall{0.75}} & \cellcolor{oursblue}\shortstack{\textbf{61.87}\sdsmall{0.15}\\\textbf{53.71}\sdsmall{0.07}}\\[3pt]
Llama-Nemotron-Embed-1B-v2 & \shortstack{66.31\\58.13} & \shortstack{66.30\sdsmall{0.22}\\57.81\sdsmall{0.15}} & \shortstack{66.03\sdsmall{0.20}\\57.87\sdsmall{0.35}} & \shortstack{66.00\sdsmall{0.12}\\57.89\sdsmall{0.03}} & \shortstack{66.18\sdsmall{0.44}\\57.76\sdsmall{0.23}} & \cellcolor{oursblue}\shortstack{\textbf{67.14}\sdsmall{0.12}\\\textbf{58.56}\sdsmall{0.21}}\\[3pt]
Llama-NV-Embed-Reasoning-3B & \shortstack{67.52\\57.23} & \shortstack{67.03\sdsmall{0.56}\\56.82\sdsmall{0.38}} & \shortstack{67.18\sdsmall{0.48}\\57.32\sdsmall{0.22}} & \shortstack{66.74\sdsmall{0.47}\\56.87\sdsmall{0.55}} & \shortstack{65.31\sdsmall{0.85}\\55.62\sdsmall{1.04}} & \cellcolor{oursblue}\shortstack{\textbf{68.35}\sdsmall{0.31}\\\textbf{58.49}\sdsmall{0.20}}\\[3pt]
Llama-Embed-Nemotron-8B & \shortstack{68.03\\59.49} & \shortstack{68.18\sdsmall{0.14}\\59.21\sdsmall{0.06}} & \shortstack{68.01\sdsmall{0.24}\\59.37\sdsmall{0.02}} & \shortstack{68.17\sdsmall{0.21}\\59.40\sdsmall{0.13}} & \shortstack{67.80\sdsmall{1.00}\\59.64\sdsmall{0.70}} & \cellcolor{oursblue}\shortstack{\textbf{69.82}\sdsmall{0.11}\\\textbf{60.61}\sdsmall{0.11}}\\[3pt]
\bottomrule\end{tabular*}\par\endgroup

\end{table}

\section{Additional mechanism controls}
\label{app:mechanisms}
All controls retain the depth, residual normalization, and fusion weight selected for NEAR. Shuffled-reference results are averaged over three fixed permutations.

\section{Efficiency details}
\label{app:cost}
Storage and latency are measured on representative memory banks using one CPU core. Storage includes the index and query-side state; query latency includes vector preparation and ANN search, excluding encoding.

\noindent\textbf{Offline construction.}
Offline construction starts from cached embeddings. We report the median of five runs.

\begin{table}[H]
\centering
\caption{\textbf{Offline bank construction (seconds).} LoCoMo and LongMemEval banks contain 629 and 476 memories, respectively.}
\begingroup\nearUnifiedTableStyle\neutralSummaryTableStyle
\begin{tabularx}{\linewidth}{@{\hspace{\nearTableEdgePadding}}>{\columncolor{white}[\nearTableEdgePadding][\tabcolsep]}l*{3}{>{\centering\arraybackslash}X}>{\columncolor{white}[\tabcolsep][\nearTableEdgePadding]\centering\arraybackslash}X@{\hspace{\nearTableEdgePadding}}}\toprule
& \multicolumn{2}{c}{\textbf{Qwen3-Embedding-0.6B}} & \multicolumn{2}{c}{\textbf{Qwen3-Embedding-4B}}\\
\cmidrule(lr){2-3}\cmidrule(l){4-5}
\textbf{Method} & LoCoMo & LongMemEval & LoCoMo & LongMemEval\\\midrule
Base (FP32) & 0.063 & 0.040 & 0.140 & 0.087\\
Base (SQ4) & 0.131 & 0.089 & 0.270 & 0.187\\
LC & 0.072 & 0.065 & 0.229 & 0.102\\
InvGC & 0.071 & 0.050 & 0.158 & 0.110\\
DBSN & 0.073 & 0.050 & 0.157 & 0.101\\
Search-Adaptor & 0.065 & 0.045 & 0.141 & 0.097\\
\rowcolor{nearTableTint}\textbf{NEAR (FP32)} & 0.166 & 0.129 & 1.147 & 0.397\\
\rowcolor{nearTableTint}\textbf{NEAR (SQ4)} & 0.274 & 0.244 & 1.988 & 0.648\\
\bottomrule\end{tabularx}\par\endgroup

\end{table}

\begin{table}[H]
\centering
\caption{\textbf{Qwen3-Embedding-0.6B quality and retrieval costs.} Quality uses complete tasks; costs use representative banks. Aux. denotes query-side state.}
\begingroup\nearUnifiedTableStyle\fontsize{8}{9.4}\selectfont\setlength{\tabcolsep}{3pt}
\begin{tabularx}{\linewidth}{@{\hspace{\nearTableEdgePadding}}>{\columncolor{white}[\nearTableEdgePadding][\tabcolsep]}l*{5}{>{\raggedleft\arraybackslash}X}>{\columncolor{white}[\tabcolsep][\nearTableEdgePadding]\raggedleft\arraybackslash}X@{\hspace{\nearTableEdgePadding}}}\toprule
\textbf{Method} & R@10 & N@10 & Index (MiB) & Aux. (KiB) & Total (MiB) & Query (ms)\\
\midrule
\rowcolor{nearGroupTint}\multicolumn{7}{c}{\textbf{LoCoMo; 629 candidates; 8 timed queries}}\\
Base (FP32) & 46.44 & 33.92 & 2.621 & 0.03 & 2.621 & 0.148\\
Base (SQ4) & 45.96 & 33.57 & 0.479 & 0.03 & 0.479 & 0.156\\
LC & 50.66 & 37.76 & 2.623 & 0.04 & 2.623 & 0.131\\
InvGC & 48.78 & 35.62 & 2.621 & 0.04 & 2.621 & 0.109\\
DBSN & 46.63 & 34.40 & 2.623 & 0.04 & 2.623 & 0.137\\
Search-Adaptor & 52.23 & 38.08 & 2.621 & 2054.34 & 4.627 & 0.267\\
\rowcolor{neutralTableGray}\textbf{NEAR (FP32)} & 55.09 & 41.28 & 5.080 & 128.82 & 5.206 & 0.347\\
\rowcolor{neutralTableGray}\textbf{NEAR (SQ4)} & 55.23 & 41.19 & 0.799 & 128.82 & 0.925 & 0.441\\
\midrule
\rowcolor{nearGroupTint}\multicolumn{7}{c}{\textbf{LongMemEval; 476 candidates; 1 timed query}}\\
Base (FP32) & 84.84 & 72.82 & 1.983 & 0.03 & 1.983 & 0.139\\
Base (SQ4) & 84.81 & 73.00 & 0.364 & 0.03 & 0.364 & 0.148\\
LC & 86.08 & 73.95 & 1.985 & 0.04 & 1.985 & 0.129\\
InvGC & 88.74 & 77.07 & 1.983 & 0.04 & 1.983 & 0.112\\
DBSN & 86.67 & 76.17 & 1.985 & 0.04 & 1.985 & 0.130\\
Search-Adaptor & 88.71 & 77.60 & 1.983 & 2054.34 & 3.990 & 0.255\\
\rowcolor{neutralTableGray}\textbf{NEAR (FP32)} & 86.45 & 75.53 & 3.843 & 128.82 & 3.969 & 0.315\\
\rowcolor{neutralTableGray}\textbf{NEAR (SQ4)} & 86.65 & 75.49 & 0.605 & 128.82 & 0.730 & 0.393\\
\bottomrule\end{tabularx}\endgroup

\end{table}
\begin{table}[H]
\centering
\caption{\textbf{Qwen3-Embedding-4B quality and retrieval costs.} Same settings as the preceding table.}
\begingroup\nearUnifiedTableStyle\fontsize{8}{9.4}\selectfont\setlength{\tabcolsep}{3pt}
\begin{tabularx}{\linewidth}{@{\hspace{\nearTableEdgePadding}}>{\columncolor{white}[\nearTableEdgePadding][\tabcolsep]}l*{5}{>{\raggedleft\arraybackslash}X}>{\columncolor{white}[\tabcolsep][\nearTableEdgePadding]\raggedleft\arraybackslash}X@{\hspace{\nearTableEdgePadding}}}\toprule
\textbf{Method} & R@10 & N@10 & Index (MiB) & Aux. (KiB) & Total (MiB) & Query (ms)\\
\midrule
\rowcolor{nearGroupTint}\multicolumn{7}{c}{\textbf{LoCoMo; 629 candidates; 8 timed queries}}\\
Base (FP32) & 41.39 & 29.69 & 6.306 & 0.03 & 6.307 & 0.396\\
Base (SQ4) & 41.38 & 29.62 & 0.951 & 0.03 & 0.951 & 0.358\\
LC & 45.12 & 32.23 & 6.309 & 0.04 & 6.309 & 0.281\\
InvGC & 49.34 & 35.88 & 6.306 & 0.04 & 6.307 & 0.319\\
DBSN & 51.93 & 38.24 & 6.309 & 0.04 & 6.309 & 0.351\\
Search-Adaptor & 50.67 & 37.81 & 6.306 & 5132.34 & 11.319 & 0.679\\
\rowcolor{neutralTableGray}\textbf{NEAR (FP32)} & 56.09 & 41.88 & 24.737 & 962.04 & 25.676 & 1.817\\
\rowcolor{neutralTableGray}\textbf{NEAR (SQ4)} & 56.38 & 41.98 & 3.318 & 962.04 & 4.258 & 1.569\\
\midrule
\rowcolor{nearGroupTint}\multicolumn{7}{c}{\textbf{LongMemEval; 476 candidates; 1 timed query}}\\
Base (FP32) & 58.96 & 47.11 & 4.773 & 0.03 & 4.773 & 0.335\\
Base (SQ4) & 59.07 & 47.09 & 0.725 & 0.03 & 0.725 & 0.322\\
LC & 58.96 & 47.11 & 4.774 & 0.04 & 4.774 & 0.219\\
InvGC & 91.52 & 81.05 & 4.773 & 0.04 & 4.773 & 0.240\\
DBSN & 91.92 & 80.56 & 4.774 & 0.04 & 4.774 & 0.280\\
Search-Adaptor & 93.28 & 82.86 & 4.773 & 5132.34 & 9.785 & 0.512\\
\rowcolor{neutralTableGray}\textbf{NEAR (FP32)} & 89.67 & 78.15 & 9.421 & 320.82 & 9.734 & 0.810\\
\rowcolor{neutralTableGray}\textbf{NEAR (SQ4)} & 89.63 & 77.97 & 1.325 & 320.82 & 1.639 & 0.740\\
\bottomrule\end{tabularx}\endgroup

\end{table}
\FloatBarrier

\end{document}